\documentclass[reprint, amsmath,amssymb, superscriptaddress,floatfix]{revtex4-2}
\pdfoutput=1
\usepackage[subpreambles=true]{standalone}
\usepackage{import, inputenc, float} 

\usepackage{amsmath, amssymb,amsfonts, caption, graphicx, epstopdf, amsthm, hyperref, subcaption}
\usepackage[dvipsnames]{xcolor}

\usepackage{fancyhdr,cancel,verbatim,listings} 	\usepackage{diagbox,float}
\usepackage[perpage]{footmisc}

\usepackage[normalem]{ulem} 
\usepackage{braket,tensor, tikz,multirow, bm} 									
\usetikzlibrary{decorations.pathmorphing}

\usepackage[space]{grffile} 		
\usepackage{marvosym} 

\usepackage{pgfplots, pgfplotstable}
\usepackage{setspace}

\usepackage{array}	
\usepackage{indentfirst}
\usepackage[titletoc]{appendix}

\usetikzlibrary{intersections, arrows,decorations.pathmorphing,snakes, shapes, positioning, patterns, decorations.text,decorations.pathreplacing}
\usepackage{tikz-3dplot}

\usetikzlibrary{shapes.gates.logic.US, calc}
\usepackage[siunitx]{circuitikz}\ctikzset{resistor = european}

\pgfdeclaredecoration{fiber}{initial}
{
    \state{initial}[width=1.5\pgfdecorationsegmentamplitude, next state=final]{
    \pgfpathmoveto{\pgfpointorigin}
    \pgflineto{\pgfpoint{\pgfdecoratedinputsegmentlength/2}{0pt}}
    \pgfpathcircle{\pgfpoint{\pgfdecoratedinputsegmentlength/2}{.5\pgfdecorationsegmentamplitude}}{.5\pgfdecorationsegmentamplitude}
    \pgfpathcircle{\pgfpoint{\pgfdecoratedinputsegmentlength/2-.25\pgfdecorationsegmentamplitude}{.5\pgfdecorationsegmentamplitude}}{.5\pgfdecorationsegmentamplitude}
    \pgfpathcircle{\pgfpoint{\pgfdecoratedinputsegmentlength/2+.25\pgfdecorationsegmentamplitude}{.5\pgfdecorationsegmentamplitude}}{.5\pgfdecorationsegmentamplitude}
    \pgfpathmoveto{\pgfpoint{\pgfdecoratedinputsegmentlength/2}{0pt}}
    }
    \state{final}{
        \pgfpathlineto{\pgfpointdecoratedpathlast}
    }
}

\newcommand{\specialcell}[2][c]{%
  \begin{tabular}[#1]{@{}c@{}}#2\end{tabular}}
\definecolor{Darkblue}{HTML}{00356b} 

\definecolor{bluelight}{rgb}{.51, 0, .78}
\definecolor{darkgreen}{rgb}{0.28, 0.81, 0}
\definecolor{neonpink}{rgb}{1, 0.23, .58}
\definecolor{neongreen}{rgb}{.502, .988, .016}
\definecolor{neoncyan}{rgb}{.33, 1, .88}
\definecolor{blue2}{rgb}{0,.353, 1}	
\definecolor{orange2}{rgb}{1, .31 0}	
\definecolor{red2}{rgb}{1, .149 0}	

\DeclareTextCompositeCommand{\r}{OT1}{A}{%
  \leavevmode\vbox{%
    \offinterlineskip
    \ialign{\hfil##\hfil\cr\char23\cr\noalign{\kern-1.15ex}A\cr}%
  }%
}

\begin{document}
\title{EIT spectroscopy of high-lying Rydberg states in $^{39}K$}

\begin{abstract}
We present a study of the Rydberg spectrum in $^{39}K$ for $nS$ and $nD_{3/2}$ series connected to $5^2P_{1/2}$ using two-photon spectroscopy based on EIT in a heated vapor cell. We observed some 80 transitions from $5P_{1/2}$ to Rydberg states with principal quantum numbers $n\sim50-90$, and determined their transition frequencies and state energies with sub-GHz precision. Our spectroscopy results lay the groundwork for using Rydberg atoms as sensitive microwave photon detectors in searches for dark matter axions in the  $\sim40-200$ $\mu$eV mass range, which is a prime range for future axion searches suggested by theory studies. 

\end{abstract}

\author{Yuqi Zhu}
\email{yuqi.zhu@yale.edu}
\affiliation{Wright Laboratory, Department of Physics, Yale University, New Haven, Connecticut 06520, USA}

\author{Sumita Ghosh}
\affiliation{Wright Laboratory, Department of Physics, Yale University, New Haven, Connecticut 06520, USA}
\affiliation{Department of Applied Physics, Yale University, New Haven, Connecticut 06520, USA}

\author{S. B. Cahn}
\affiliation{Wright Laboratory, Department of Physics, Yale University, New Haven, Connecticut 06520, USA}

\author{M. J.\ Jewell}
\affiliation{Wright Laboratory, Department of Physics, Yale University, New Haven, Connecticut 06520, USA}
\author{D. H.\ Speller}
\affiliation{Department of Physics, Johns Hopkins University, Baltimore, Maryland 21218, USA}

\author{Reina H.\ Maruyama}
\email{reina.maruyama@yale.edu}
\affiliation{Wright Laboratory, Department of Physics, Yale University, New Haven, Connecticut 06520, USA}

\date{\today}
\maketitle

\section{Introduction}
The axion is a well-motivated solution to the strong charge–parity ($CP$) problem in quantum chromodynamics (QCD), and also a natural dark matter candidate \cite{Peccei, Peccei2, Weinberg1978, Wilczek1978}. Depending on the cosmological history, a wide range of axion masses, $m_a \sim 10^{-6}-10^3\:\mu$eV, can provide the correct abundance of dark matter \cite{marsh20161}. Due to this wide range and the extremely weak coupling to matter required by benchmark QCD models \cite{Kim1979, DINE1981}, the axion remains undetected. 
Among the ongoing axion searches, the most sensitive searches are resonant microwave cavity experiments, 
covering the axion mass range $m_a \sim 0.2-25$ $\mu$eV \cite{ADMX2018, ADMX2020, ADMX2021, brubaker2017first, HAYSTAC2018, HAYSTAC2021}. They make use of the axion-to-photon conversion in a strong magnetic field and enhance the conversion by tuning the cavity's resonance to the hypothetical mass \cite{Sikivie1983}. Given the typical parameters in an ongoing microwave cavity experiment \cite{HAYSTAC2021}, there are $\sim0.3$ photons per second, or equivalently $\sim 10^{-24}$ W at $m_a= 19$ $\mu$eV, converted from axions of the Milky Way halo \cite{Sikivie1985}. Therefore, making sensitive and low-noise measurements of the photon number is critical for axion detection.

This study is motivated by the potential of using potassium atoms in Rydberg states as sensitive microwave probes in a resonant cavity search for dark matter axions \cite{tada1999, Yamamoto2001}. Rydberg atoms are well suited to single-photon detection and microwave electrometry for having strong coupling to electromagnetic fields and long radiative lifetimes. In single-photon detection via absorption of the axion converted photon, the primary source of measurement background and noise is blackbody photons. At cryogenic temperatures, single-photon detectors are more compelling than linear amplifiers for searches at higher masses, $m_a>40 \: \mu\text{eV } (\approx 10$ GHz) \cite{Lamoreaux2013}. 
Previously $^{85}$Rb Rydberg atoms have been considered for this application \cite{TADA2006, tada1999}. Compared to rubidium, dipole transitions between low-angular-momentum states in $^{39}$K Rydberg atoms are less sensitive to the dc Stark effect for having similar polarizabilities. Therefore single-photon detection would be less sensitive to stray electric fields.
To enhance the sensitivity to axion, a suitable detection scheme can be chosen by matching the resonant frequency of a dipole transition between Rydberg states to the target axion mass. 
For instance, to search at $m_a\approx 40$ $\mu$eV, the most sensitive Rydberg states---with dipole transitions in the vicinity---would have principal quantum numbers $n \sim 90$. 
According to some recent calculations, $m_a\sim 40-200\: \mu$eV ($\approx 10-50$ GHz) is a highly probable mass range \cite{Ballesteros2017, Gorghetto2021, buschmann2021}. High-lying Rydberg states with $n\sim60-90$ in $^{39}$K would be particularly relevant to future axion searches in this mass range (see Appendix for further details). However, only a few of the Rydberg states with $n>50$ have been identified in previous spectroscopy studies \cite{Lorenzen1981, Lorenzen1983, Thompson1983, Chen2020}. Therefore, we study the energy spectrum of high-lying Rydberg states with $n\sim50-90$ using an all-optical detection based on electromagnetically induced transparency (EIT) \cite{Boller1991, Mohapatra2007, Xu2016}. Broadly, beyond single-photon detection, Rydberg and Rydberg-dressed atoms have attracted interest as tools for quantum  many-body physics and quantum information \cite{Saffman2010, Pupillo2010}. High-lying Rydberg states have longer lifetime compared to low-lying states, as the lifetime scales as $n^3$ for the low-angular-momentum case. 

In the remainder of this paper, we describe the experiment, including the Rydberg excitation scheme, experimental setup and analysis procedure, in Sec.~\ref{sec: exp}. Then in Sec.~\ref{sec: data}, we present the spectroscopy data with comparisons to other literature values, and then the derivation of quantum defects and ionization energies from data. In Sec.~\ref{sec: summary}, we give a summary and outlook.

\section{Experiment \label{sec: exp}}
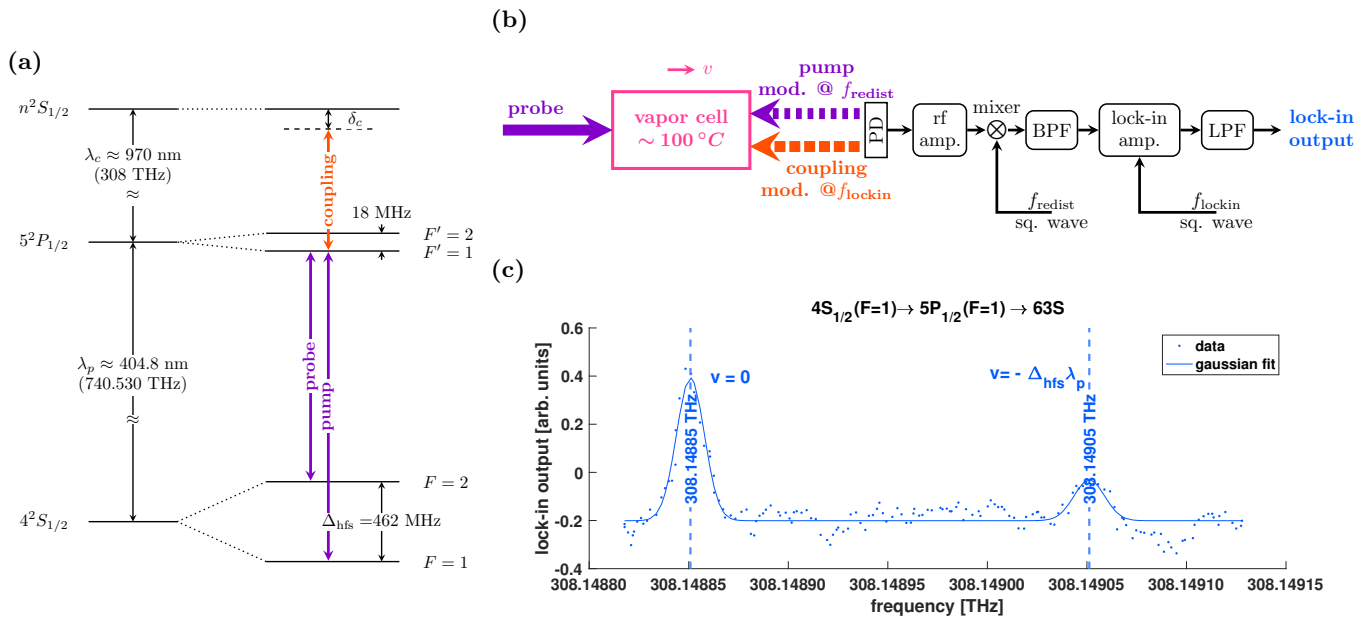
\begin{figure*}\centering
\begin{subfigure}{.35\textwidth} \centering
   \caption{} \label{fig: EITlevels}
  \resizebox{\textwidth}{!}{
\begin{tikzpicture}[scale=1.8, level/.style={thick}, trans/.style={thick,<->, >=stealth}, laser/.style={ultra thick, <->, >=stealth, shorten >=0.5, shorten <=0.5, orange2},  laserb/.style={ultra thick, <->, >=stealth, shorten >=0.5, shorten <=0.5, bluelight}, p/.style = {midway, fill=white, inner sep=1}
  ]
    \pgfmathsetmacro\s{-.5} 
    \pgfmathsetmacro\p{3} 
     \pgfmathsetmacro\ss{4.5} 

    \pgfmathsetmacro\h{.3} 
    \pgfmathsetmacro\g{-.6} 
    \pgfmathsetmacro\f{1} 
    \pgfmathsetmacro\ww{.5} 
    \pgfmathsetmacro\w{1.5} 

  \node at (\s,\g/2+\h/2){4$^2S_{1/2}$ };
  \node at (\s,\p){5$^2P_{1/2}$};
  \node at (\s,\ss){$n^2S_{1/2}$};

  \draw[level] (0,\g/2+\h/2)--(1,\g/2+\h/2) (0,\p)--(1,\p) (0,\ss)--(1,\ss);
  \draw[level](2,\g)--(2+\w,\g) node[right, xshift=1em]{$F=1$};
  \draw[level](2,\h)--(2+\w,\h)node[right, xshift=1em]{$F=2$};
\draw[level](2,\ss)--(2+\w,\ss)node[right, xshift=1em]{};

 \draw[thick,dotted] (2,\g)--(1,\g/2+\h/2)--(2,\h);
 \draw[thick,dotted] (2,\ss)--(1,\ss);

   \draw[trans](1-.5,\p)--(1-.5,\g/2+\h/2);
  \node[fill=white,inner sep=.1] at (.5, \p/2){\specialcell{$\lambda_p\approx$ 404.8 nm\\ (740.530 THz)}};
  \node[fill=white,inner sep=.1] at (.5, \p/2-.5){$\approx$};
  
    \draw[trans](1-.5,\ss)--(1-.5,\p) node[midway,fill=white,inner sep=.1]{\specialcell{$\lambda_c \approx$ 970 nm\\ (308 THz) \\ $\approx$}};

\draw[level](2,\p-0.1*\f)--(2+\w,\p-0.1*\f)node[right, xshift=1em]{$F'=1$};
\draw[level](2,\p+0.1*\f)--(2+\w,\p+0.1*\f)node[right, xshift=1em]{$F'=2$ };
\draw[thick, dotted] (2,\p-0.1*\f)--(1,\p)--(2,\p+0.1*\f);

\pgfmathsetmacro\pp{2.5} %
\draw[laserb] (\pp,\h)--(\pp,\p-0.1*\f) node[p, sloped] {\bf probe};
\draw[laser] (\pp+.2,\p-0.1*\f) --(\pp+.2,\ss*.95)node[p, sloped] {\bf coupling};
\draw[laserb] (\pp+.2,\g)--(\pp+.2,\p-0.1*\f) node[p, sloped] {\bf pump};
\draw[level, dashed](\pp-.3,\ss*.95)--++(1,0);
\draw[trans] (\pp+.2,\ss*.95) --(\pp+.2,\ss)node[p, right, xshift=1em] {\bf $\delta_c$};
 \draw[trans,<-](2+\w-.2,\p-0.1*\f)--++(0,-.1); \draw[trans,<-](2+\w-.2,\p+0.1*\f)--++(0,.1) node[above]{18 MHz};
  \draw[trans](2+\w-.2,\g)--(2+\w-.2,\h) node[p]{$\Delta_\text{hfs}=$462 MHz};

\end{tikzpicture}}
\end{subfigure}%
~
 \begin{subfigure}{.65\textwidth} \centering
 \caption{\label{fig: EITsetup}}
 \resizebox{\textwidth}{!}{
\begin{tikzpicture}[scale=1, octagon/.style=
  {shape=regular polygon, regular polygon sides=4, draw, minimum width=4cm}, beam/.style={->, line width=2 mm, orange2,  >=stealth},  signal/.style={->, line width=.5 mm,  >=stealth},box/.style ={rectangle, draw=black, thick, text width=3.5em, align=center, rounded corners, minimum height=2em}]
  
   \pgfmathsetmacro \c{9.5} 
\pgfmathsetmacro \p{3} 
 \pgfmathsetmacro \l{2}
    \pgfmathsetmacro \pd{10} 

\draw[ultra thick, neonpink] (5,.7) rectangle (7.5,-.7) node [midway]{\specialcell{\bf vapor cell \\\bm{$\sim 100 \: {}^\circ C$}}}; 

\draw[signal, neonpink] (6,1.1)--++(.5,0) node[right]{\bf $v$}; 

\draw[beam, dotted]  (\c,-.3)--++(-\l,0) node[pos=.3, below] {\bf \specialcell{ coupling\\ mod. @$f_\text{lockin} $}};
\draw[beam, bluelight, dashed]  (\c, 0.3)--++(-\l,0) node[pos=.3, above] {\bf \specialcell{pump\\ mod. @ $f_\text{redist}$}};
\draw[beam, bluelight]  (\p,0)--++(\l,0) node[pos=.3, above] {\bf probe};

\draw[thick] (\c+.1,.5) rectangle ( \pd,-.5)  node[pos=.5,rotate=90] {PD}; 
\draw ( \pd,0) node[left] (pda) {} 
( \pd+.48*\l,0) node[box,,text width=2.1em] (amp) {rf amp.} 
(\pd+1*\l,0) node[ inner sep=0em] (mixer) {$\bigotimes$}
(\pd+1.5*\l,0) node[box,text width=2em] (bpf){BPF}
(\pd+2.3*\l, 0) node[box] (lockinamp) {lock-in amp.}
(\pd+3.1*\l,0) node[box,text width=2em] (lpf){LPF};

\node[above] at (mixer.north)  {mixer};
\draw[signal] (pda.east)--(amp.west);
\draw[signal](amp.east)--(mixer.west);
\draw[signal] (\pd+1.5*\l,-1.5) node []{\specialcell{$f_\text{redist}$\\ sq. wave}} -|(mixer.south);
\draw[signal](mixer.east)--(bpf.west);
\draw[signal](bpf.east)--(lockinamp.west);
\draw[signal](lockinamp.east)--(lpf.west);
\draw[signal] (\pd+3.*\l,-1.5) node []{\specialcell{$f_\text{lockin}$\\ sq. wave}} -|(lockinamp.south);
\draw[signal] (lpf.east)--++(.5,0) node[right, blue2] {\specialcell{\bf lock-in\\\bf output}};
\end{tikzpicture}
}  
\caption{ \label{fig: ex}}
\includegraphics[width=.9\textwidth]{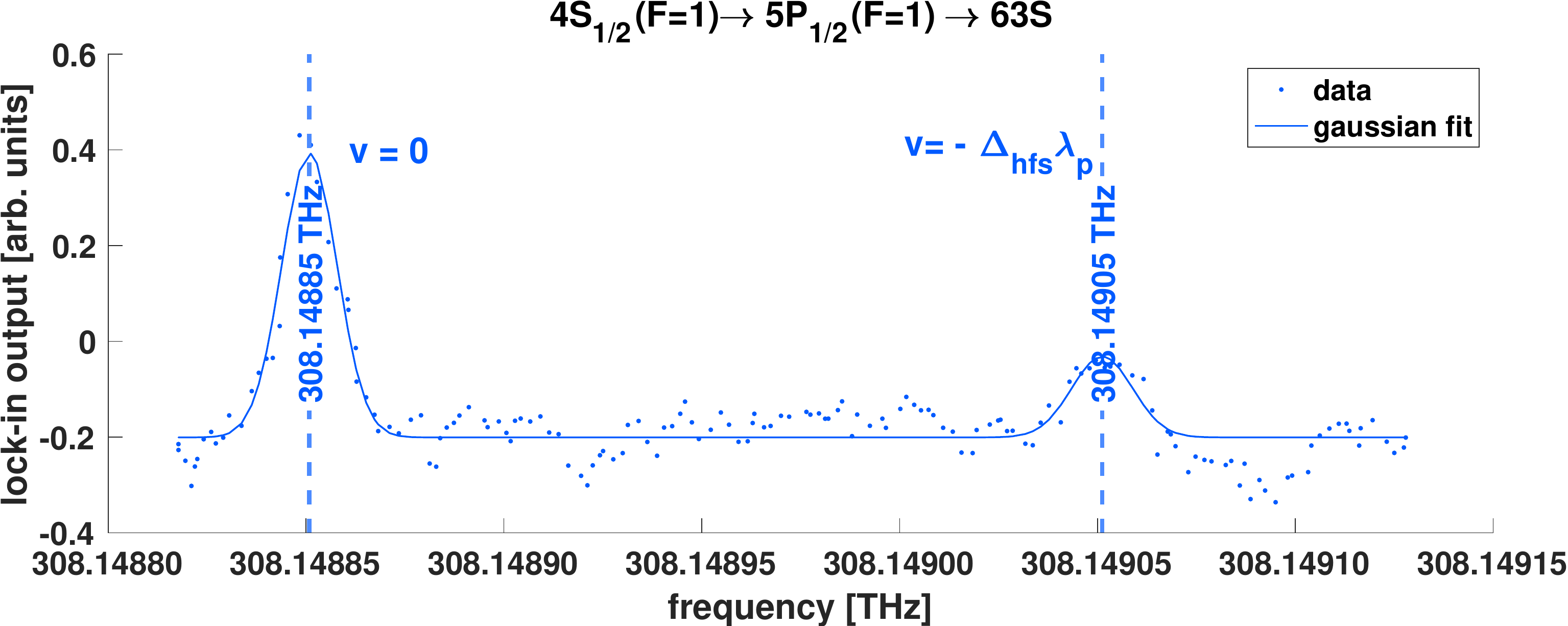}
 \end{subfigure}
\caption{(a) Energy levels and transitions in EIT. In EIT spectroscopy, the probe and pump frequencies are fixed while the frequency of the coupling beam is scanned to identify transitions from $5^2P_{1/2}$ to Rydberg states. When coupling frequency is on resonance, i.e., the coupling detuning $\delta_c=0$, maximum probe transmission is attained. (b) Experimental setup. All three beams are aligned to overlap with each other in a vapor cell heated to $\sim 100^\circ$C. The intensities of the coupling and pump beams are modulated at different frequencies for lock-in detection ($f_\text{lockin}$) and ground state population redistribution ($f_\text{redist}$) respectively. The lock-in output is produced by doubly demodulating the amplified transmission signal. (c) An EIT spectrum near the $5P_{1/2}-63S$ Rydberg transition. Each data point is averaged over 10 samples, and the peak locations, indicated by dashed lines, are found by fitting the spectrum to a sum of two gaussian functions.} 
\end{figure*}
In this experiment, the Rydberg excitation involves two resonant transitions connected to the $5^2P_{1/2}$ intermediate state: the first transition, $4S-5P_{1/2}$, is at about 405 nm, and the second Rydberg transition, $5P_{1/2}-\{nS, nD_{3/2}\}$, at about 970 nm for $n\sim 50-90$. Alternatively, one can use one of the $4P$ states as the intermediate state for Rydberg excitation. The transitions between the intermediate state ($4P$ or $5P$) and Rydberg states are usually weak. For instance, the dipole matrix elements of transitions between $5P_{1/2}$ and Rydberg levels with $n \sim50-90$ are on the order of 0.01 $ea_0$, which is about 100 times weaker than the $D_2$ line used in laser cooling and 10 times weaker than the $4S-5P_1/2$ transition. Therefore, the former scheme involving $5P$ presents a technological advantage---the small dipole matrix element of the transition between a Rydberg state and the intermediate $5P$ state can be matched by powerful infrared laser. On a more fundamental level, however, the low-high wavelength ordering makes observing EIT challenging. Consider an atom moving at velocity $v$ in the lab frame along the propagation direction of the probe beam that drives the $4S-5P_{1/2}$ transition. Because the probe wavelength ($\lambda_p=2\pi/k_p \approx 405$ nm) is shorter than the coupling wavelength ($\lambda_c=2\pi/k_c\approx $ 970 nm), the one-photon Doppler shift, $- k_p v$, has the same sign as the two-photon Doppler shift, $-(k_p v - k_c v)$. Consequently, the transparency window as a function of the coupling detuning no longer exists, and the EIT feature in Doppler-averaged absorption is smaller by many orders of magnitude than that in a system where the probe wavelength is longer than the coupling wavelength \cite{Boon1999, Urvoy2013}. 

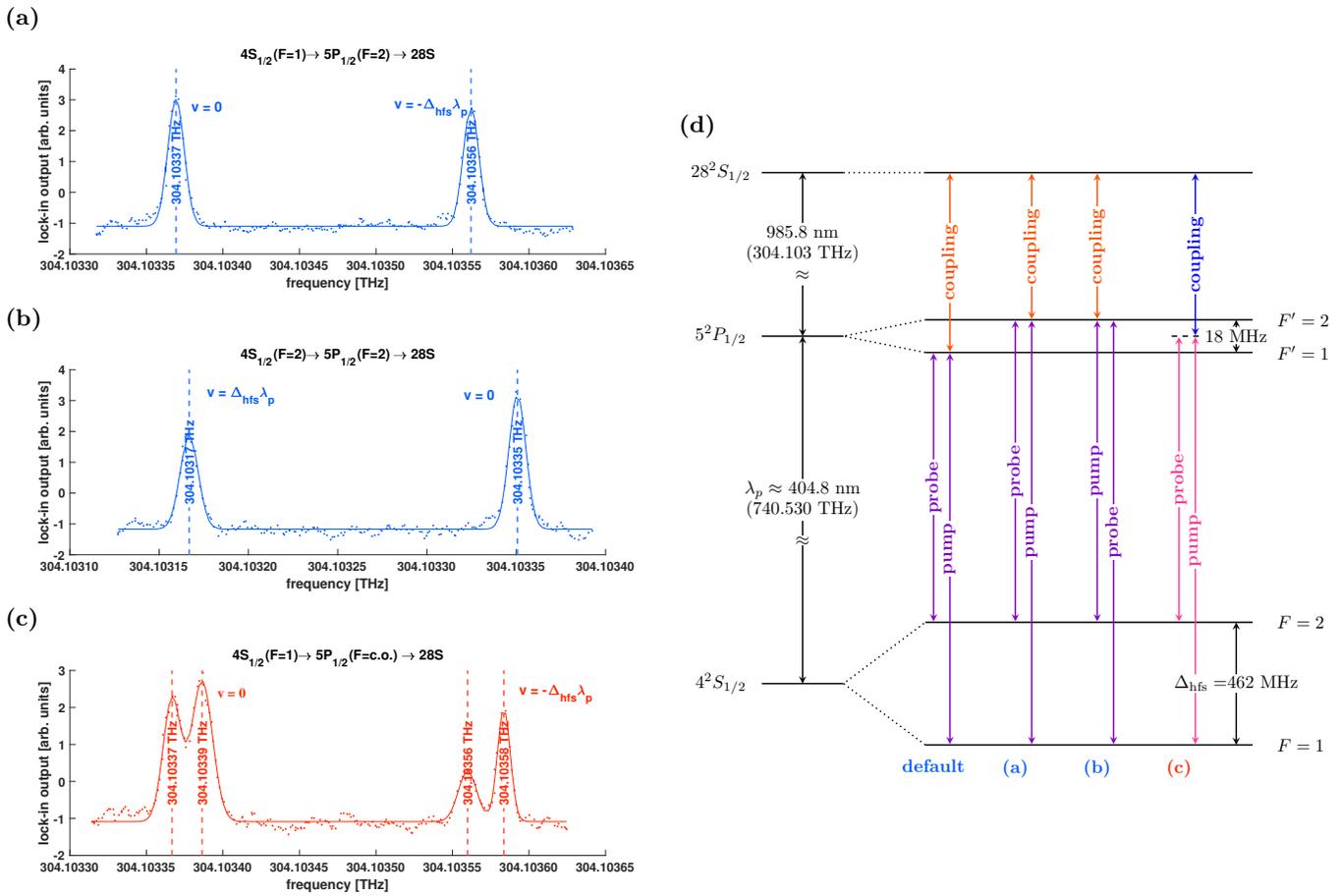
\begin{figure*}\centering
  \begin{subfigure}{.5\textwidth} \centering
    \caption{}
    \includegraphics[width=.9\textwidth]{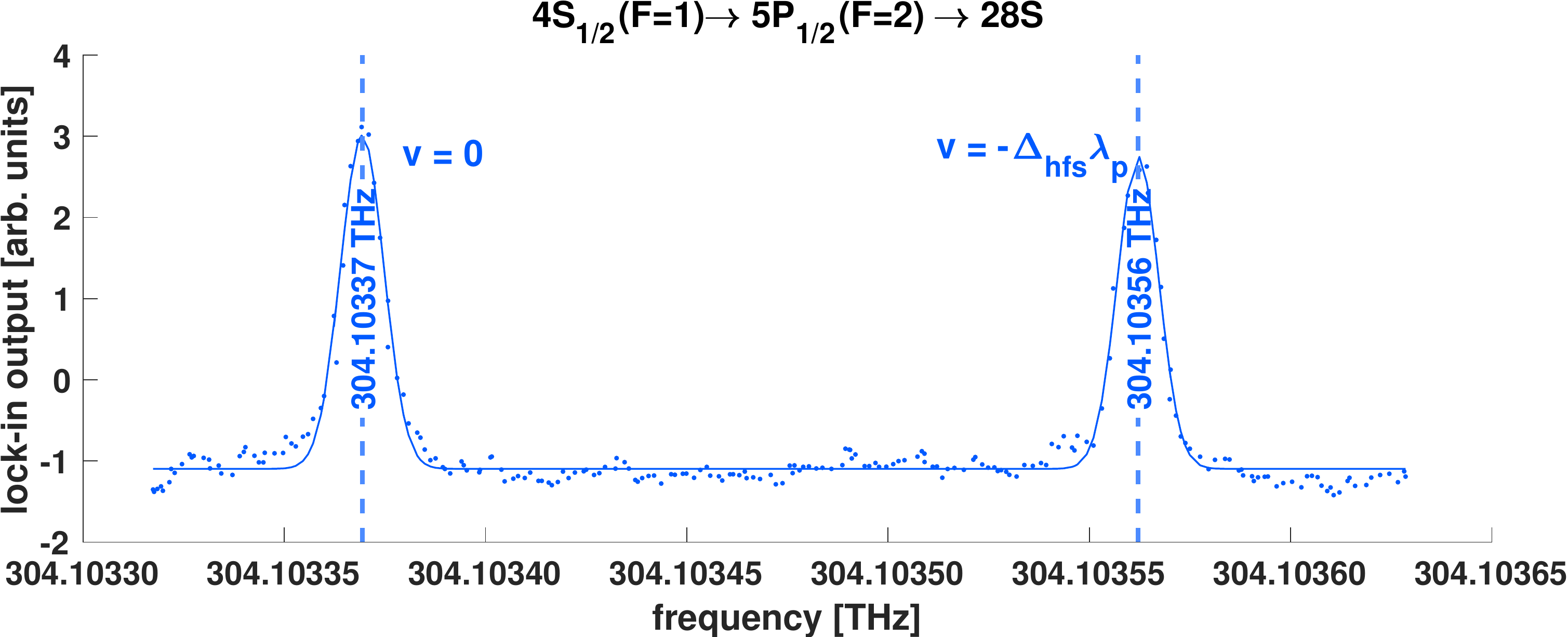}
    \caption{}
    \includegraphics[width=.9\textwidth]{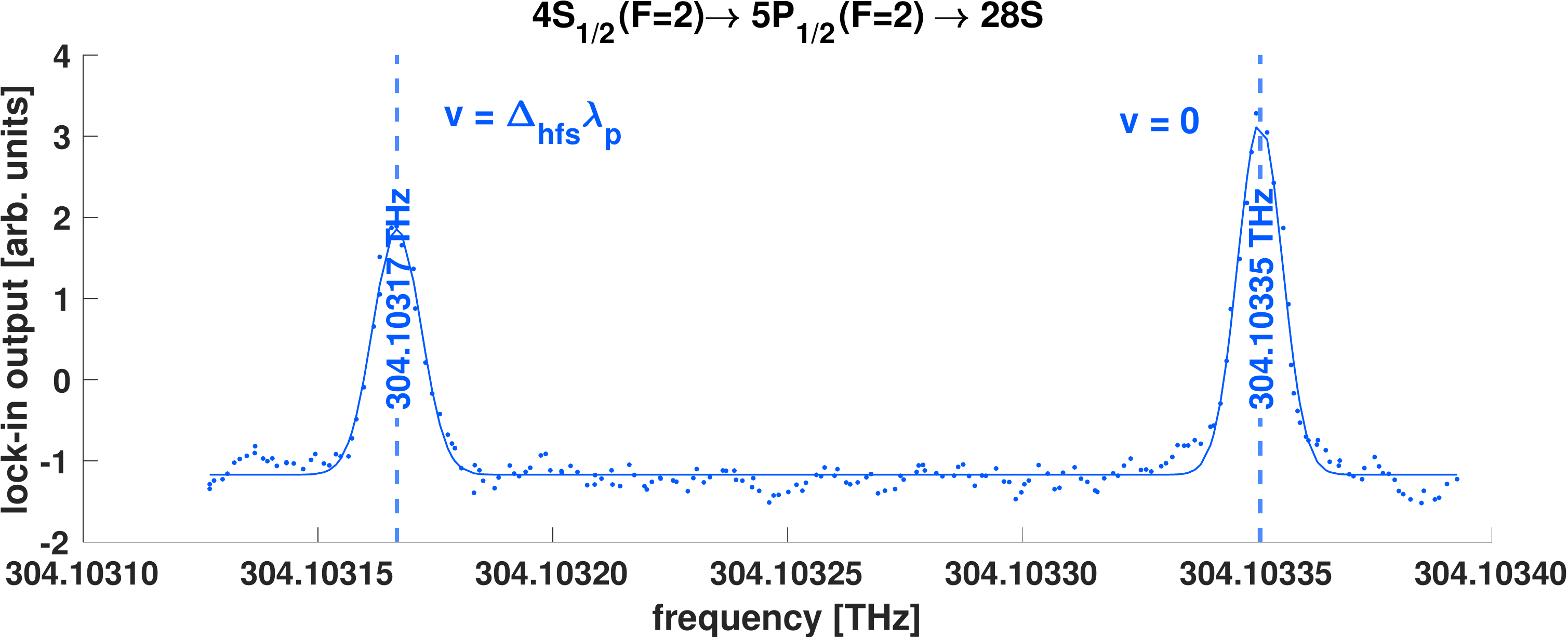}
     \caption{}
    \includegraphics[width=.9\textwidth]{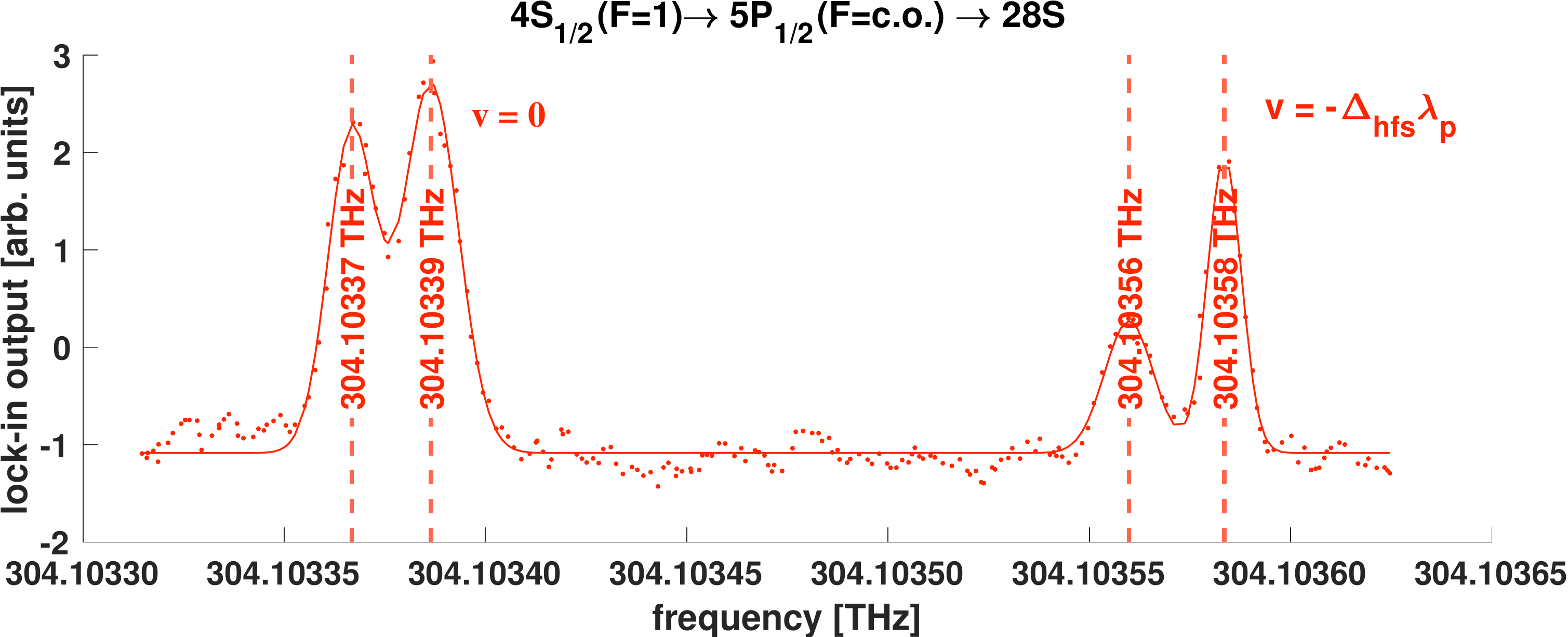}
     \end{subfigure}%
   ~
    \begin{subfigure}{.5\textwidth} \centering
    \caption{}
    \resizebox{\textwidth}{!}{
\begin{tikzpicture}[scale=1.6, level/.style={thick}, trans/.style={thick,<->, >=stealth}, laser/.style={thick, <->, >=stealth, shorten >=0.5, shorten <=0.5, orange2},  laserb/.style={thick, <->, >=stealth, shorten >=0.5, shorten <=0.5, bluelight}, p/.style = {midway, fill=white, inner sep=1}
  ]
    \pgfmathsetmacro\s{-.5} 
    \pgfmathsetmacro\p{4} 
     \pgfmathsetmacro\ss{6} 

    \pgfmathsetmacro\h{.5} 
    \pgfmathsetmacro\g{-1} 
    \pgfmathsetmacro\f{2} 
    \pgfmathsetmacro\w{4} 

  \node at (\s,\g/2+\h/2){4$^2S_{1/2}$ };
  \node at (\s,\p){5$^2P_{1/2}$};
    \node at (\s,\ss){28$^2S_{1/2}$};

  \draw[level] (0,\g/2+\h/2)--(1,\g/2+\h/2) (0,\p)--(1,\p) (0,\ss)--(1,\ss);
  \draw[level](2,\g)--(2+\w,\g) node[right, xshift=1em]{$F=1$};
  \draw[level](2,\h)--(2+\w,\h)node[right, xshift=1em]{$F=2$};
\draw[level](2,\ss)--(2+\w,\ss)node[right, xshift=1em]{};

  \draw[thick,dotted] (2,\g)--(1,\g/2+\h/2)--(2,\h);
 \draw[thick,dotted] (2,\ss)--(1,\ss);
 
    \draw[trans](1-.5,\p)--(1-.5,\g/2+\h/2);
  \node[fill=white,inner sep=.1] at (.5, \p/2){\specialcell{$\lambda_p\approx$ 404.8 nm\\ (740.530 THz)}};
  \node[fill=white,inner sep=.1] at (.5, \p/2-.5){$\approx$};
    \draw[trans](1-.5,\ss)--(1-.5,\p) node[midway,fill=white,inner sep=.1]{\specialcell{985.8 nm\\ (304.103 THz) \\ $\approx$}};

\draw[level](2,\p-0.1*\f)--(2+\w,\p-0.1*\f)node[right, xshift=1em]{$F'=1$};
\draw[level](2,\p+0.1*\f)--(2+\w,\p+0.1*\f)node[right, xshift=1em]{$F'=2$ };
\draw[thick, dotted] (2,\p-0.1*\f)--(1,\p)--(2,\p+0.1*\f);

\pgfmathsetmacro\pp{2.1} 

\draw[laserb] (\pp,\h)--(\pp,\p-0.1*\f) node[p, sloped] {\bf probe};
\draw[laser] (\pp+.2,\p-0.1*\f) --(\pp+.2,\ss)node[p, sloped] {\bf coupling};
\draw[laserb] (\pp+.2,\g)--(\pp+.2,\p-0.1*\f) node[p, sloped] {\bf pump};
\node[below, blue2 ] at (\pp, \g-.1){\bf default};

\draw[laserb] (\pp+1,\h)--(\pp+1,\p+0.1*\f) node[p, sloped] {\bf probe};
\draw[laser] (\pp+1+.2,\p+0.1*\f) --(\pp+1+.2,\ss)node[p, sloped] {\bf coupling};
\draw[laserb] (\pp+1+.2,\g)--(\pp+1+.2,\p+0.1*\f) node[p, sloped] {\bf pump};
\node[below, blue2 ] at (\pp+1, \g-.1) {\bf (a)};

\draw[laserb] (\pp+2,\h)--(\pp+2,\p+0.1*\f) node[p, sloped] {\bf pump};
\draw[laser] (\pp+2,\p+0.1*\f) --(\pp+2,\ss)node[p, sloped] {\bf coupling};
\draw[laserb] (\pp+.2+2,\g)--(\pp+.2+2,\p+0.1*\f) node[p, sloped] {\bf probe};
\node[below, blue2 ] at (\pp+2, \g-.1) {\bf (b)};

\draw[laser, neonpink] (\pp+3,\h)--(\pp+3,\p) node[p, sloped] {\bf probe};
\draw[laser, blue] (\pp+3+.2,\p) --(\pp+3+.2,\ss)node[p, sloped] {\bf coupling};
\draw[laser,neonpink] (\pp+3+.2,\g)--(\pp+3+.2,\p) node[p, sloped] {\bf pump};
\draw[level, dashed](2+\w,\p)--++(-1,0);
\node[below, red2] at (\pp+3, \g-.1) {\bf (c)};

  \draw[trans](2+\w-.2,\p+0.1*\f)--(2+\w-.2,\p-0.1*\f) node[p]{18 MHz};
  \draw[trans](2+\w-.2,\g)--(2+\w-.2,\h) node[p]{$\Delta_\text{hfs}=$462 MHz};
\end{tikzpicture}    }
    \end{subfigure}
  \caption{Optical pumping in EIT. (a-b) The pump beam drives a resonance transition between two hyperfine levels.  (c) The pump beam drives a cross-over transition on resonance. (d) EIT schemes used in (a-c). }\label{fig: op}
\end{figure*}

To mitigate the suppression of EIT, we implement the velocity-selective EIT scheme presented in Ref.~\cite{Xu2016}, which combines lock-in detection with an optical pumping scheme that eliminates any dark states in 4S. Fig.~\ref{fig: EITlevels} shows the relevant energy levels and transitions. The pump and probe are tuned to be on resonance with $4S(F=1)-5P_{1/2}(F'=1)$ and $4S(F=2)-5P_{1/2}(F'=1)$ respectively. They come from the same external cavity diode laser (ECDL, model: TOPTICA DL PRO) and the frequency of the pump beam is offset by the ground state hyperfine splitting $\Delta_\text{hfs}$ using a double-pass acousto-optic modulator setup. The frequency of the ECDL is locked using a Doppler-free saturated absorption signal as the reference. The resulting uncertainty is less than 9 MHz---half the hyperfine splitting of $5P_{1/2}$. The probe and pump deliver about 2 mW and respectively 10 mW to the experiment, and their diameters are roughly equal, $\sim$ 2 mm. The coupling beam provides $50-60$ mW depending on the specific wavelength, and its diameter is $\sim .5$ mm. Fig.~\ref{fig: EITsetup} shows the experimental setup. To look for the small EIT feature in the probe transmission measured by a photodiode (PD), the lock-in technique is applied---the lock-in modulation is added by chopping the coupling beam at $f_\text{lockin}\approx 1.3 $ kHz. In addition to modulating the intensity of the coupling beam, the intensity of the pump beam is also modulated with 100\% modulation intensity by a square wave in order to reduce the population imbalance between the two hyperfine ground states due to strong optical pumping on the 405-nm transition. The redistribution frequency $f_\text{redist}=25$ kHz is chosen to be slower than the interatomic collision rate, which is about 50 kHz given the cell temperature and the 2-mm beam size \cite{Urvoy2013}. This effectively modulates the population hole and peak near $v=0$ at $f_\text{redist}$. The transmission signal is amplified by the PD and an additional radio frequency (rf) amplifier (amp.) before doubly demodulated by mixing first with a square wave of frequency $f_\text{redist}$ at an rf mixer and then with a square wave of frequency $f_\text{lockin}$ in a commercial lock-in amplifier (Stanford Research 510) with built-in input and output filters. A bandpass filter (BPF) is applied to the input signal; it centers at $f_\text{lockin}$ and has a 6-dB roll off in either direction. A low-pass filter (LPF) is applied to the output after the demodulation; it has a cutoff frequency of $.53$ Hz and provides 6 dB/oct attenuation. The lock-in output is eventually used to determine the Rydberg transition frequencies.

Fig.~\ref{fig: ex} shows an example of an EIT spectrum near the $5P_{1/2}-28S$ transition. The horizontal axis shows the frequency of the coupling beam measured by a wavemeter (HighFinesse WS/7) whose absolute accuracy and resolution are 60 MHz and 1 MHz respectively.  
Because the pump and probe frequency are offset by $\Delta_\text{hfs}$, two velocity groups are observed in the EIT spectrum. For the EIT scheme in Fig.~\ref{fig: EITlevels}, they are $v=0$ and $v=-\Delta_\text{hfs} \lambda_p$. By convention, $v$, the velocity in the lab frame, is positive if it is parallel to the arrow above the cell in Figure~\ref{fig: EITsetup}. The $v=-\Delta_\text{hfs}\lambda_p$ atoms absorb blue-shifted photons from the probe beam and red-shifted photons from the coupling beam. Thus the maximum transmission, also the most efficient Rydberg excitation for this velocity group, occurs at a higher frequency than that for $v=0$ atoms that absorb from the pump and coupling beams. 
We note that our spectrum consists of two peaks instead of four as in Ref.\cite{Xu2016}. Comparing different optical pumping schemes in EIT (Fig.~\ref{fig: op}), we see that when the pump is on resonance with a transition between energy eigenstates, only one intermediate hyperfine level is populated. The locations of the $v=0$ peaks are consistent in different spectra, and the separation between the $v=0$ and $v=-\Delta_\text{hfs} \lambda_p$ peaks is $\Delta_\text{hfs}\frac{\lambda_p}{\lambda_c} \approx 190$ MHz in all cases.

\section{Results \label{sec: data}}
All observed Rydberg levels, $nS$ and $nD_{3/2}$, can be found in Table~\ref{tab: transitions}, where $f_c$ is the measured coupling frequency from $5P_{1/2}(F'=1)$ to the Rydberg state and  $E_{n,l,j}$ is the measured state energy relative to the $4S(F=1)$ hyperfine ground state. Compared to other experimentally measured energies of $^{39}$K Rydberg states (Table~\ref{tab: compareE}), ours agree with the most recent results from Ref.\cite{Chen2020} within uncertainties (except for $68S$) but differ from earlier results in Refs.\cite{Lorenzen1981, Thompson1983}. The observed transition frequency of the first transition, $4S-5P_{1/2}$, is not consistent with the NIST value by 570(60) MHz \cite{Lorenzen1981, Sansonetti} but consistent with the reported value in Ref.\cite{Xu2016}.

\begin{figure} \centering
\includegraphics[width=\columnwidth]{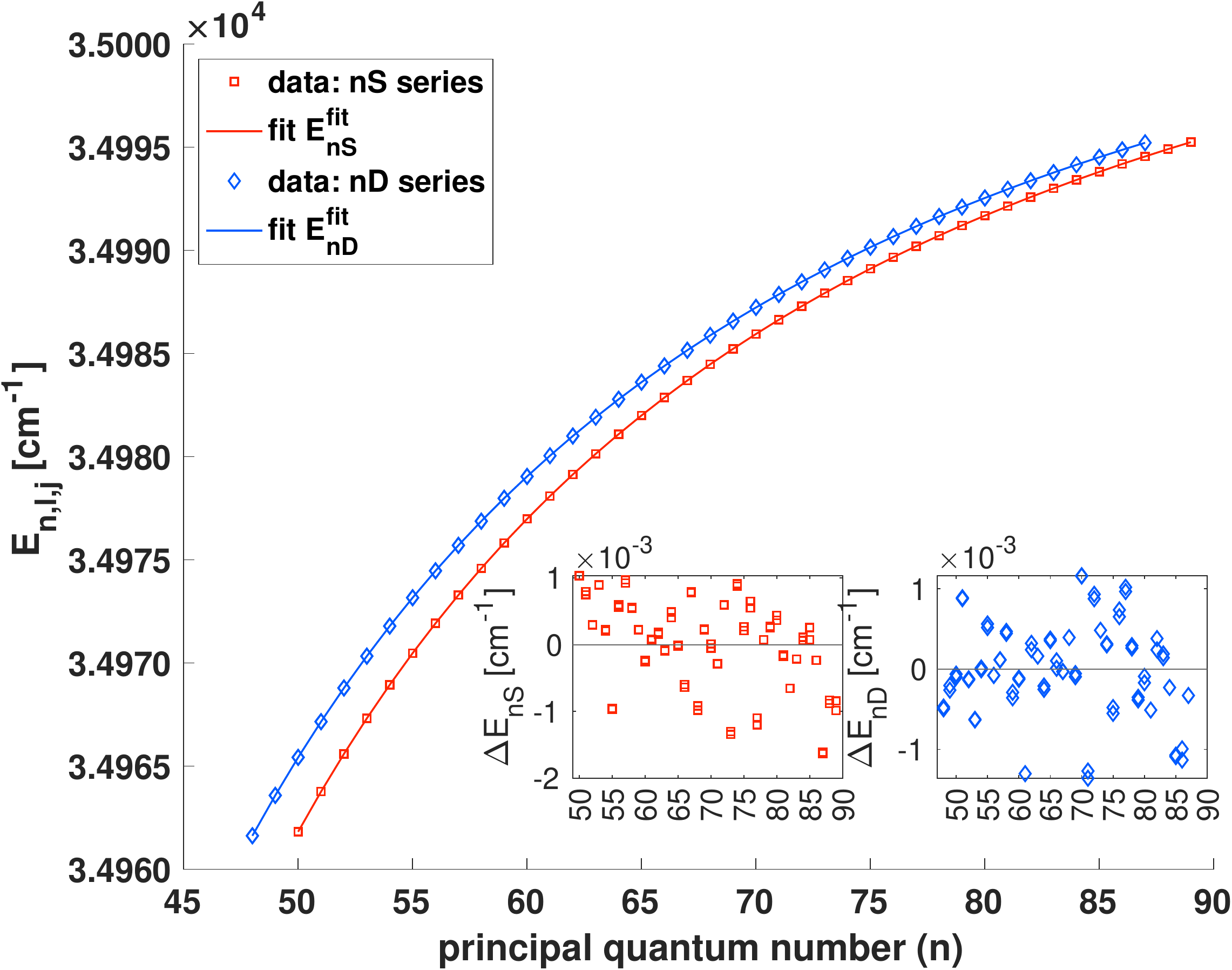}
\caption{State energies fitted to the Rydberg energy equation with two Rydberg-Ritz coefficients. Insets show the difference between data and fit for each series $\Delta E=E-E^\text{fit}$. The adjust residual squared ($R^2$) deviates from 1 by $10^{-14}$ and $10^{-9}$ for $nS$ and $nD_{3/2}$ series respectively.  }        \label{fig: Rylevels}
\end{figure}

From measured state energies, the ionization energy, $E_\infty$, and the quantum defect, $\delta(n, l, j)$, can be determined by fitting to \cite{gallagher1994}
\begin{align}
    E_{n,l, j} &= E_\infty -\frac{\text{Ry}^*}{(n-\delta(n,l,j))^2},\label{eq: RyE}
\end{align}
where Ry$^*= 109 735.770 665 6$ cm$^{-1}$ is the Rydberg constant for $^{39}$K, numerically calculated from fundamental constants \cite{NISTCODATA, AtomicWeight, Peper2019}. 
The quantum defect in \eqref{eq: RyE} has a phenomenological expression,
\begin{equation}
    \delta(n,l,j) =\delta_0 + \frac{\delta_2 }{(n-\delta_0 )^2} +o(n^4),
\end{equation}
where $\{\delta_{2i}\}_{i=0}$ are the Rydberg-Ritz coefficients specific to an angular momentum configuration \cite{gallagher1994, Martin1980}. The quantum defect most strongly depends on $l$, the orbital angular momentum, as it is caused by the penetration of the electron orbital into the ionic core. Writing it as a series expansion is in the same spirit of adding spherically symmetric corrections to the Coulomb potential in perturbation theory \cite{Drake1991}. 
Fig.~\ref{fig: Rylevels} shows a plot of fit results for both term series. The ionization energy is 35009.805(3) cm$^{-1}$ for both series. The quantum defects, to leading order and independent of $n$, are $\delta_0=2.181(1)$ for the $nS$ states and $\delta_0=0.277(1)$ for $nD_{3/2}$.  Statistical uncertainties from fitting are negligible, and the quoted uncertainties are systematic propagated from the wavemeter's accuracy. 
Our data of high-lying Rydberg states is not very sensitive to high order terms in \eqref{eq: RyE}. Notwithstanding this, our $\delta_0$ agrees with results derived using similar methods in Refs.\cite{Sansonetti, Lorenzen1981,  Lorenzen1983, Thompson1983, Chen2020, Peper2019}. Our ionization energy is consistent with the reported value in Ref.\cite{Chen2020}, but it appears to be lower than most other reported values \cite{Sansonetti, Lorenzen1981,  Lorenzen1983, Thompson1983, Peper2019}. 

\begin{table}
\caption{Comparing observed state energies relative to the $4S(F=1)$ ground state to other experimental results. The uncertainty in our values is systematic, due to the wavemeter, 80 MHz $\approx 0.003$ cm$^{-1}$. }
\label{tab: compareE}
\resizebox{\columnwidth}{!}{
 \begin{tabular}{l c c c  c c} \hline\hline
 & \multicolumn{4}{c}{state energy $E_{n,l,j}$ [cm$^{-1}$]}\\
excited state &  observed & NIST \cite{NIST_ASD, Sansonetti} & Ref.\cite{Thompson1983} & Ref.\cite{Chen2020}  \\
  \hline
$5P_{1/2}$ & 24701.399(2) & 24701.38(3) \\
28S        & 34845.196(3)  & 34845.2158(7)  & 34845.214(3) & 34845.18(3)  \\
50S       & 34961.817(3)  &                & 34961.835(3) &  34961.80(3) \\
54S       & 34968.939(3)  &                &              & 34968.94(3)\\
55S       &  34970.470(3)  &                & 34970.493(3)    & \\
56S       &  34971.920(3)  &                &  & 34971.89(3)\\
62S       & 34979.139(3) & & &             34979.13(3) \\
66S     & 34982.862(3) & & & 34982.83(3)\\
68S       & 34984.474(3) &&                & 34984.36(3)\\
70S & 34985.947(3) & & & 34985.93(3) \\
\hline\hline
\end{tabular}
    }
\end{table}

\begin{table*}\centering
\caption{Observed coupling frequencies ($f_c$) and energies ($E_{n,l,j}$) of Rydberg states. State energies are calculated by adding the measured transition energy of $4S(F=1)-5P_{1/2}(F=1)$, 24701.399(2) cm$^{-1}$, to the coupling frequencies. The uncertainties (not shown) are systematic, 60 MHz for $f_c$ and 0.003 cm$^{-1}$ for $E_{n,l,j}$. }
    \label{tab: transitions}
\begin{subtable}{0.3\textwidth}
\resizebox{\textwidth}{!}{
 \begin{tabular}{l c c c  c c} \hline\hline
Ry. state  & $f_c$ [THz] & $E_{n,0,1/2}$ [cm$^{-1}$] & \\
  \hline
28S &304.10338 &34845.196\\
50S &307.59958 &34961.817\\
51S &307.65791 &34963.762\\
52S &307.71275 &34965.592\\
53S &307.76442 &34967.315\\
54S &307.81309 &34968.939\\
55S &307.85901 &34970.470\\
56S &307.90247 &34971.920\\
57S &307.94354 &34973.290\\
58S &307.98240 &34974.586\\
59S &308.01923 &34975.815\\
61S &308.08734 &34978.087\\
62S &308.11887 &34979.139\\
63S &308.14885 &34980.139\\
64S &308.17741 &34981.091\\
65S &308.20458 &34981.998\\
66S &308.23049 &34982.862\\
67S &308.25526 &34983.688\\
68S &308.27882 &34984.474\\
69S &308.30141 &34985.227\\
70S &308.32297 &34985.947\\
71S &308.34360 &34986.635\\
72S &308.36339 &34987.295\\
73S &308.38225 &34987.924\\
74S &308.40046 &34988.531\\
75S &308.41784 &34989.111\\
76S &308.43454 &34989.668\\
77S &308.45052 &34990.201\\
78S &308.46596 &34990.716\\
79S &308.48077 &34991.210\\
80S &308.49500 &34991.685\\
81S &308.50868 &34992.141\\
82S &308.52186 &34992.581\\
83S &308.53457 &34993.005\\
84S &308.54681 &34993.413\\
85S &308.55861 &34993.807\\
86S &308.56998 &34994.186\\
87S &308.58091 &34994.550\\
88S &308.59153 &34994.905\\
89S &308.60176 &34995.246\\
 \hline\hline
\end{tabular}
}
\end{subtable}%
~
\begin{subtable}{0.3\textwidth}
\resizebox{\textwidth}{!}{
 \begin{tabular}{l c c c  c c} \hline\hline
Ry. state  & $f_c$ [THz] & $E_{n,2, 3/2}$ [cm$^{-1}$] & \\
  \hline
48$D_{3/2}$ &307.59373 &34961.622\\
49$D_{3/2}$ &307.65242 &34963.579\\
50$D_{3/2}$ &307.70760 &34965.420\\
51$D_{3/2}$ &307.75958 &34967.154\\
52$D_{3/2}$ &307.80852 &34968.786\\
53$D_{3/2}$ &307.85471 &34970.327\\
54$D_{3/2}$ &307.89837 &34971.784\\
55$D_{3/2}$ &307.93967 &34973.161\\
56$D_{3/2}$ &307.97873 &34974.464\\
57$D_{3/2}$ &308.01576 &34975.699\\
58$D_{3/2}$ &308.05089 &34976.871\\
59$D_{3/2}$ &308.08421 &34977.982\\
60$D_{3/2}$ &308.11589 &34979.039\\
61$D_{3/2}$ &308.14598 &34980.043\\
62$D_{3/2}$ &308.17471 &34981.001\\
63$D_{3/2}$ &308.20202 & 34981.912\\
64$D_{3/2}$ &308.22805 &34982.780\\
65$D_{3/2}$ &308.25291 &34983.609\\
66$D_{3/2}$ &308.27661 &34984.400\\
67$D_{3/2}$ &308.29927 &34985.156\\
68$D_{3/2}$ &308.32094 &34985.879\\
69$D_{3/2}$ &308.34165 &34986.570\\
70$D_{3/2}$ &308.36152 &34987.233\\
71$D_{3/2}$ &308.38045 &34987.864\\
72$D_{3/2}$ &308.39873 &34988.474\\
73$D_{3/2}$ &308.41619 &34989.056\\
74$D_{3/2}$ &308.43294 &34989.615\\
75$D_{3/2}$ &308.44901 &34990.151\\
76$D_{3/2}$ &308.46451 &34990.668\\
77$D_{3/2}$ &308.47937 &34991.163\\
78$D_{3/2}$ &308.49364 &34991.639\\
79$D_{3/2}$ &308.50737 &34992.097\\
80$D_{3/2}$ &308.52061 &34992.539\\
81$D_{3/2}$ &308.53334 &34992.964\\
82$D_{3/2}$ &308.54564 &34993.374\\
83$D_{3/2}$ &308.55748 &34993.769\\
84$D_{3/2}$ &308.56888 &34994.149\\
85$D_{3/2}$ &308.57987 &34994.516\\
86$D_{3/2}$ &308.59050 &34994.870\\
87$D_{3/2}$ &308.60079 &34995.214\\
\hline\hline
\end{tabular}
}
\end{subtable}
\end{table*}

\section{Summary and Outlook \label{sec: summary}}
In summary, we directly measured the transition frequencies between $5P_{1/2}$ and 81 Rydberg states with $n\sim 50-90$ using two-photon spectroscopy based on velocity-selective EIT, and calculated the their state energies. Out of the 81 Rydberg states we found, 9 have been previously identified in experiments; the state energies determined using our method are in agreement with most recent experimental results. 

As searches for the QCD axion extend to higher masses, specifically for $m_a > 40$ $\mu$eV, single-photon detection will become increasingly relevant. 
This work marks a step towards developing single-photon detection using Rydberg atoms for this application. Being able to populate high-$n$ low-angular-momentum Rydberg states via two-photon transitions is fundamental to preparing atoms into other detection states, such as circular Rydberg states which are Rydberg states the maximum possible orbital angular momentum, $l$, projection along the quantization axis ($l=n-1=|m_l|$). The Doppler-free EIT spectrum will also enable us to stabilize the frequency of the 970-nm laser driving the Rydberg transition to good precision and accuracy. Furthermore, the spectroscopy data may also be relevant to future applications of Rydberg and Rydberg-dressed atoms for quantum many-body physics and quantum information.

\begin{acknowledgments}
We would like to thank Mark Saffman and Michael Peper for fruitful discussions. We would also like to thank Gabe Hoshino, Elizabeth Ruddy, Annie Giman, and Sophia Getz for support on the experiment. Y. Z., S. G., S. C., M. J., and R. M. are supported in part by Department of Energy under grant number DE-AC02-07CH11359;  S. G. is supported in part by the National Science Foundation under grant number DMR-1747426. D. S. would like to thank Johns Hopkins University. 
\end{acknowledgments}

\appendix*
\section{Rydberg atom-based microwave detection \label{sec: ap}}
\begin{figure}
    \centering
    \includegraphics[width=\columnwidth]{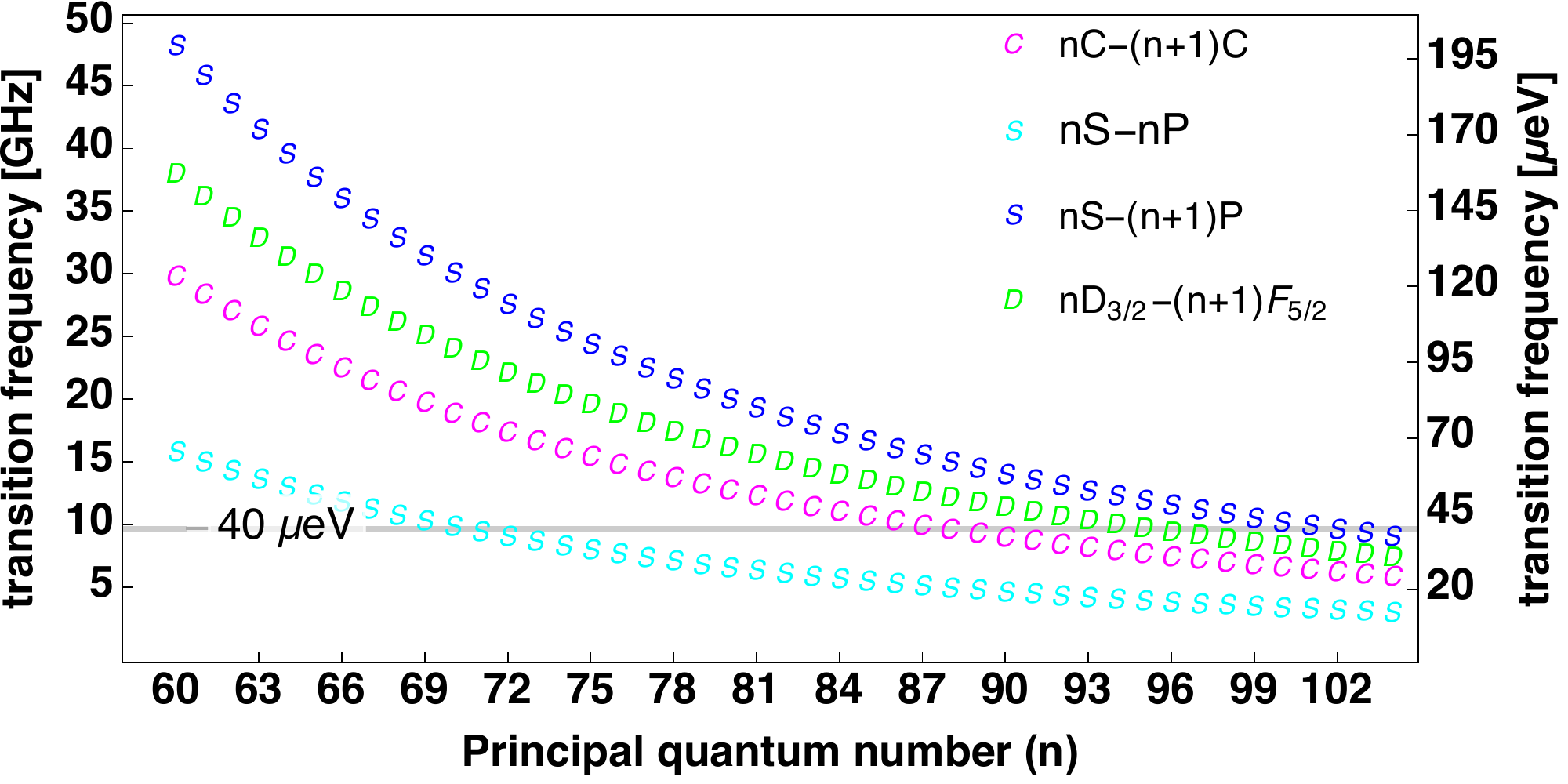}
    \caption{Transition frequency vs principal quantum number. The $y-$axes are equivalent and only different in units. The most sensitive Rydberg states for $m_a= 40 \:\mu$eV $\approx$ 10 GHz are $70S$ (or $101S$), $87C$, and $95D_{3/2}$. The fine structure splitting within $nP$ is on the order of $0.5\:\mu$eV. However, the resulting difference in transition frequency is not resolved in this plot. }
    \label{fig: transitions}
\end{figure}

In this appendix, we discuss the prospect of microwave detection using Rydberg atoms in the context of an axion search to clarify why Rydberg states with $n\sim60-90$ are particularly relevant to axion searches in the $m_a\sim 40-200 \: \mu$eV range. Suppose, as in CARRACK II, the detection mechanism is a combination of microwave absorption and selective field ionization \cite{Yamamoto2001}. 
Since the frequency of the axion-converted photon we wish to detect is equal to $m_a$, a suitable transition for a given axion mass target should be as close to the axion mass as possible to increase the probability of absorption. There are potentially many suitable transitions connecting different initial and final states. Their transition frequencies can be calculated using the Alkali Rydberg Calculator (ARC) \cite{ARC}; using the default parameters in ARC, quantum defects and ionization energy used to determine state energies are based on Ref.\cite{Lorenzen1983}. 
Fig.~\ref{fig: transitions} shows the transition frequency vs principal quantum number for dipole transitions starting from three types of initial states. $nS$ and $nD_{3/2}$ are readily accessible via the two-photon excitation scheme used in this work. $nC$ are representative of high-$l$ states, which have much longer lifetimes and larger transition dipole moments compared to low-$l$ states. Preparing them require an additional transferring step, known as the adiabatic rapid passage (ARP) \cite{Signoles2017}, after the two-photon optical excitation.

\bibliography{Ryspectroscopy}

\begin{thebibliography}{43}%
\makeatletter
\providecommand \@ifxundefined [1]{%
 \@ifx{#1\undefined}
}%
\providecommand \@ifnum [1]{%
 \ifnum #1\expandafter \@firstoftwo
 \else \expandafter \@secondoftwo
 \fi
}%
\providecommand \@ifx [1]{%
 \ifx #1\expandafter \@firstoftwo
 \else \expandafter \@secondoftwo
 \fi
}%
\providecommand \natexlab [1]{#1}%
\providecommand \enquote  [1]{``#1''}%
\providecommand \bibnamefont  [1]{#1}%
\providecommand \bibfnamefont [1]{#1}%
\providecommand \citenamefont [1]{#1}%
\providecommand \href@noop [0]{\@secondoftwo}%
\providecommand \href [0]{\begingroup \@sanitize@url \@href}%
\providecommand \@href[1]{\@@startlink{#1}\@@href}%
\providecommand \@@href[1]{\endgroup#1\@@endlink}%
\providecommand \@sanitize@url [0]{\catcode `\\12\catcode `\$12\catcode
  `\&12\catcode `\#12\catcode `\^12\catcode `\_12\catcode `\%12\relax}%
\providecommand \@@startlink[1]{}%
\providecommand \@@endlink[0]{}%
\providecommand \url  [0]{\begingroup\@sanitize@url \@url }%
\providecommand \@url [1]{\endgroup\@href {#1}{\urlprefix }}%
\providecommand \urlprefix  [0]{URL }%
\providecommand \Eprint [0]{\href }%
\providecommand \doibase [0]{https://doi.org/}%
\providecommand \selectlanguage [0]{\@gobble}%
\providecommand \bibinfo  [0]{\@secondoftwo}%
\providecommand \bibfield  [0]{\@secondoftwo}%
\providecommand \translation [1]{[#1]}%
\providecommand \BibitemOpen [0]{}%
\providecommand \bibitemStop [0]{}%
\providecommand \bibitemNoStop [0]{.\EOS\space}%
\providecommand \EOS [0]{\spacefactor3000\relax}%
\providecommand \BibitemShut  [1]{\csname bibitem#1\endcsname}%
\let\auto@bib@innerbib\@empty
\bibitem [{\citenamefont {Peccei}\ and\ \citenamefont
  {Quinn}(1977{\natexlab{a}})}]{Peccei}%
  \BibitemOpen
  \bibfield  {author} {\bibinfo {author} {\bibfnamefont {R.~D.}\ \bibnamefont
  {Peccei}}\ and\ \bibinfo {author} {\bibfnamefont {H.~R.}\ \bibnamefont
  {Quinn}},\ }\bibfield  {title} {\bibinfo {title} {$\mathrm{CP}$ conservation
  in the presence of pseudoparticles},\ }\href
  {https://doi.org/10.1103/PhysRevLett.38.1440} {\bibfield  {journal} {\bibinfo
   {journal} {Phys. Rev. Lett.}\ }\textbf {\bibinfo {volume} {38}},\ \bibinfo
  {pages} {1440} (\bibinfo {year} {1977}{\natexlab{a}})}\BibitemShut {NoStop}%
\bibitem [{\citenamefont {Peccei}\ and\ \citenamefont
  {Quinn}(1977{\natexlab{b}})}]{Peccei2}%
  \BibitemOpen
  \bibfield  {author} {\bibinfo {author} {\bibfnamefont {R.~D.}\ \bibnamefont
  {Peccei}}\ and\ \bibinfo {author} {\bibfnamefont {H.~R.}\ \bibnamefont
  {Quinn}},\ }\bibfield  {title} {\bibinfo {title} {Constraints imposed by
  $\mathrm{CP}$ conservation in the presence of pseudoparticles},\ }\href
  {https://doi.org/10.1103/PhysRevD.16.1791} {\bibfield  {journal} {\bibinfo
  {journal} {Phys. Rev. D}\ }\textbf {\bibinfo {volume} {16}},\ \bibinfo
  {pages} {1791} (\bibinfo {year} {1977}{\natexlab{b}})}\BibitemShut {NoStop}%
\bibitem [{\citenamefont {Weinberg}(1978)}]{Weinberg1978}%
  \BibitemOpen
  \bibfield  {author} {\bibinfo {author} {\bibfnamefont {S.}~\bibnamefont
  {Weinberg}},\ }\bibfield  {title} {\bibinfo {title} {A new light boson?},\
  }\href {https://doi.org/10.1103/PhysRevLett.40.223} {\bibfield  {journal}
  {\bibinfo  {journal} {Phys. Rev. Lett.}\ }\textbf {\bibinfo {volume} {40}},\
  \bibinfo {pages} {223} (\bibinfo {year} {1978})}\BibitemShut {NoStop}%
\bibitem [{\citenamefont {Wilczek}(1978)}]{Wilczek1978}%
  \BibitemOpen
  \bibfield  {author} {\bibinfo {author} {\bibfnamefont {F.}~\bibnamefont
  {Wilczek}},\ }\bibfield  {title} {\bibinfo {title} {Problem of strong {$P$}
  and {$T$} invariance in the presence of instantons},\ }\href
  {https://doi.org/10.1103/PhysRevLett.40.279} {\bibfield  {journal} {\bibinfo
  {journal} {Phys. Rev. Lett.}\ }\textbf {\bibinfo {volume} {40}},\ \bibinfo
  {pages} {279} (\bibinfo {year} {1978})}\BibitemShut {NoStop}%
\bibitem [{\citenamefont {Marsh}(2016)}]{marsh20161}%
  \BibitemOpen
  \bibfield  {author} {\bibinfo {author} {\bibfnamefont {D.~J.}\ \bibnamefont
  {Marsh}},\ }\bibfield  {title} {\bibinfo {title} {Axion cosmology},\ }\href
  {https://doi.org/https://doi.org/10.1016/j.physrep.2016.06.005} {\bibfield
  {journal} {\bibinfo  {journal} {Physics Reports}\ }\textbf {\bibinfo {volume}
  {643}},\ \bibinfo {pages} {1} (\bibinfo {year} {2016})}\BibitemShut {NoStop}%
\bibitem [{\citenamefont {Kim}(1979)}]{Kim1979}%
  \BibitemOpen
  \bibfield  {author} {\bibinfo {author} {\bibfnamefont {J.~E.}\ \bibnamefont
  {Kim}},\ }\bibfield  {title} {\bibinfo {title} {Weak-interaction singlet and
  strong $\mathrm{CP}$ invariance},\ }\href
  {https://doi.org/10.1103/PhysRevLett.43.103} {\bibfield  {journal} {\bibinfo
  {journal} {Phys. Rev. Lett.}\ }\textbf {\bibinfo {volume} {43}},\ \bibinfo
  {pages} {103} (\bibinfo {year} {1979})}\BibitemShut {NoStop}%
\bibitem [{\citenamefont {Dine}\ \emph {et~al.}(1981)\citenamefont {Dine},
  \citenamefont {Fischler},\ and\ \citenamefont {Srednicki}}]{DINE1981}%
  \BibitemOpen
  \bibfield  {author} {\bibinfo {author} {\bibfnamefont {M.}~\bibnamefont
  {Dine}}, \bibinfo {author} {\bibfnamefont {W.}~\bibnamefont {Fischler}},\
  and\ \bibinfo {author} {\bibfnamefont {M.}~\bibnamefont {Srednicki}},\
  }\bibfield  {title} {\bibinfo {title} {A simple solution to the strong {CP}
  problem with a harmless axion},\ }\href
  {https://doi.org/https://doi.org/10.1016/0370-2693(81)90590-6} {\bibfield
  {journal} {\bibinfo  {journal} {Physics Letters B}\ }\textbf {\bibinfo
  {volume} {104}},\ \bibinfo {pages} {199} (\bibinfo {year}
  {1981})}\BibitemShut {NoStop}%
\bibitem [{\citenamefont {Du}\ \emph {et~al.}(2018)\citenamefont {Du},
  \citenamefont {Force}, \citenamefont {Khatiwada}, \citenamefont {Lentz},
  \citenamefont {Ottens}, \citenamefont {Rosenberg}, \citenamefont {Rybka},
  \citenamefont {Carosi}, \citenamefont {Woollett}, \citenamefont {Bowring}
  \emph {et~al.}}]{ADMX2018}%
  \BibitemOpen
  \bibfield  {author} {\bibinfo {author} {\bibfnamefont {N.}~\bibnamefont
  {Du}}, \bibinfo {author} {\bibfnamefont {N.}~\bibnamefont {Force}}, \bibinfo
  {author} {\bibfnamefont {R.}~\bibnamefont {Khatiwada}}, \bibinfo {author}
  {\bibfnamefont {E.}~\bibnamefont {Lentz}}, \bibinfo {author} {\bibfnamefont
  {R.}~\bibnamefont {Ottens}}, \bibinfo {author} {\bibfnamefont {L.~J.}\
  \bibnamefont {Rosenberg}}, \bibinfo {author} {\bibfnamefont {G.}~\bibnamefont
  {Rybka}}, \bibinfo {author} {\bibfnamefont {G.}~\bibnamefont {Carosi}},
  \bibinfo {author} {\bibfnamefont {N.}~\bibnamefont {Woollett}}, \bibinfo
  {author} {\bibfnamefont {D.}~\bibnamefont {Bowring}}, \emph {et~al.}
  (\bibinfo {collaboration} {ADMX Collaboration}),\ }\bibfield  {title}
  {\bibinfo {title} {Search for invisible axion dark matter with the axion dark
  matter experiment},\ }\href {https://doi.org/10.1103/PhysRevLett.120.151301}
  {\bibfield  {journal} {\bibinfo  {journal} {Phys. Rev. Lett.}\ }\textbf
  {\bibinfo {volume} {120}},\ \bibinfo {pages} {151301} (\bibinfo {year}
  {2018})}\BibitemShut {NoStop}%
\bibitem [{\citenamefont {Braine}\ \emph {et~al.}(2020)\citenamefont {Braine},
  \citenamefont {Cervantes}, \citenamefont {Crisosto}, \citenamefont {Du},
  \citenamefont {Kimes}, \citenamefont {Rosenberg}, \citenamefont {Rybka},
  \citenamefont {Yang}, \citenamefont {Bowring}, \citenamefont {Chou} \emph
  {et~al.}}]{ADMX2020}%
  \BibitemOpen
  \bibfield  {author} {\bibinfo {author} {\bibfnamefont {T.}~\bibnamefont
  {Braine}}, \bibinfo {author} {\bibfnamefont {R.}~\bibnamefont {Cervantes}},
  \bibinfo {author} {\bibfnamefont {N.}~\bibnamefont {Crisosto}}, \bibinfo
  {author} {\bibfnamefont {N.}~\bibnamefont {Du}}, \bibinfo {author}
  {\bibfnamefont {S.}~\bibnamefont {Kimes}}, \bibinfo {author} {\bibfnamefont
  {L.~J.}\ \bibnamefont {Rosenberg}}, \bibinfo {author} {\bibfnamefont
  {G.}~\bibnamefont {Rybka}}, \bibinfo {author} {\bibfnamefont
  {J.}~\bibnamefont {Yang}}, \bibinfo {author} {\bibfnamefont {D.}~\bibnamefont
  {Bowring}}, \bibinfo {author} {\bibfnamefont {A.~S.}\ \bibnamefont {Chou}},
  \emph {et~al.} (\bibinfo {collaboration} {ADMX Collaboration}),\ }\bibfield
  {title} {\bibinfo {title} {Extended search for the invisible axion with the
  axion dark matter experiment},\ }\href
  {https://doi.org/10.1103/PhysRevLett.124.101303} {\bibfield  {journal}
  {\bibinfo  {journal} {Phys. Rev. Lett.}\ }\textbf {\bibinfo {volume} {124}},\
  \bibinfo {pages} {101303} (\bibinfo {year} {2020})}\BibitemShut {NoStop}%
\bibitem [{\citenamefont {Bartram}\ \emph {et~al.}(2021)\citenamefont
  {Bartram}, \citenamefont {Braine}, \citenamefont {Burns}, \citenamefont
  {Cervantes}, \citenamefont {Crisosto}, \citenamefont {Du}, \citenamefont
  {Korandla}, \citenamefont {Leum}, \citenamefont {Mohapatra}, \citenamefont
  {Nitta} \emph {et~al.}}]{ADMX2021}%
  \BibitemOpen
  \bibfield  {author} {\bibinfo {author} {\bibfnamefont {C.}~\bibnamefont
  {Bartram}}, \bibinfo {author} {\bibfnamefont {T.}~\bibnamefont {Braine}},
  \bibinfo {author} {\bibfnamefont {E.}~\bibnamefont {Burns}}, \bibinfo
  {author} {\bibfnamefont {R.}~\bibnamefont {Cervantes}}, \bibinfo {author}
  {\bibfnamefont {N.}~\bibnamefont {Crisosto}}, \bibinfo {author}
  {\bibfnamefont {N.}~\bibnamefont {Du}}, \bibinfo {author} {\bibfnamefont
  {H.}~\bibnamefont {Korandla}}, \bibinfo {author} {\bibfnamefont
  {G.}~\bibnamefont {Leum}}, \bibinfo {author} {\bibfnamefont {P.}~\bibnamefont
  {Mohapatra}}, \bibinfo {author} {\bibfnamefont {T.}~\bibnamefont {Nitta}},
  \emph {et~al.} (\bibinfo {collaboration} {ADMX Collaboration}),\ }\href@noop
  {} {\bibinfo {title} {Search for "invisible" axion dark matter in the
  $3.3\text{-}4.2~{\mu}$ev mass range}} (\bibinfo {year} {2021}),\ \Eprint
  {https://arxiv.org/abs/2110.06096} {arXiv:2110.06096 [hep-ex]} \BibitemShut
  {NoStop}%
\bibitem [{\citenamefont {Brubaker}\ \emph {et~al.}(2017)\citenamefont
  {Brubaker}, \citenamefont {Zhong}, \citenamefont {Gurevich}, \citenamefont
  {Cahn}, \citenamefont {Lamoreaux}, \citenamefont {Simanovskaia},
  \citenamefont {Root}, \citenamefont {Lewis}, \citenamefont {Al~Kenany},
  \citenamefont {Backes} \emph {et~al.}}]{brubaker2017first}%
  \BibitemOpen
  \bibfield  {author} {\bibinfo {author} {\bibfnamefont {B.}~\bibnamefont
  {Brubaker}}, \bibinfo {author} {\bibfnamefont {L.}~\bibnamefont {Zhong}},
  \bibinfo {author} {\bibfnamefont {Y.}~\bibnamefont {Gurevich}}, \bibinfo
  {author} {\bibfnamefont {S.}~\bibnamefont {Cahn}}, \bibinfo {author}
  {\bibfnamefont {S.}~\bibnamefont {Lamoreaux}}, \bibinfo {author}
  {\bibfnamefont {M.}~\bibnamefont {Simanovskaia}}, \bibinfo {author}
  {\bibfnamefont {J.}~\bibnamefont {Root}}, \bibinfo {author} {\bibfnamefont
  {S.}~\bibnamefont {Lewis}}, \bibinfo {author} {\bibfnamefont
  {S.}~\bibnamefont {Al~Kenany}}, \bibinfo {author} {\bibfnamefont
  {K.}~\bibnamefont {Backes}}, \emph {et~al.},\ }\bibfield  {title} {\bibinfo
  {title} {First results from a microwave cavity axion search at 24
  $\mu$e{V}},\ }\href@noop {} {\bibfield  {journal} {\bibinfo  {journal} {Phys.
  Rev. Lett.}\ }\textbf {\bibinfo {volume} {118}},\ \bibinfo {pages} {061302}
  (\bibinfo {year} {2017})}\BibitemShut {NoStop}%
\bibitem [{\citenamefont {Zhong}\ \emph {et~al.}(2018)\citenamefont {Zhong},
  \citenamefont {Al~Kenany}, \citenamefont {Backes}, \citenamefont {Brubaker},
  \citenamefont {Cahn}, \citenamefont {Carosi}, \citenamefont {Gurevich},
  \citenamefont {Kindel}, \citenamefont {Lamoreaux}, \citenamefont {Lehnert}
  \emph {et~al.}}]{HAYSTAC2018}%
  \BibitemOpen
  \bibfield  {author} {\bibinfo {author} {\bibfnamefont {L.}~\bibnamefont
  {Zhong}}, \bibinfo {author} {\bibfnamefont {S.}~\bibnamefont {Al~Kenany}},
  \bibinfo {author} {\bibfnamefont {K.~M.}\ \bibnamefont {Backes}}, \bibinfo
  {author} {\bibfnamefont {B.~M.}\ \bibnamefont {Brubaker}}, \bibinfo {author}
  {\bibfnamefont {S.~B.}\ \bibnamefont {Cahn}}, \bibinfo {author}
  {\bibfnamefont {G.}~\bibnamefont {Carosi}}, \bibinfo {author} {\bibfnamefont
  {Y.~V.}\ \bibnamefont {Gurevich}}, \bibinfo {author} {\bibfnamefont {W.~F.}\
  \bibnamefont {Kindel}}, \bibinfo {author} {\bibfnamefont {S.~K.}\
  \bibnamefont {Lamoreaux}}, \bibinfo {author} {\bibfnamefont {K.~W.}\
  \bibnamefont {Lehnert}}, \emph {et~al.},\ }\bibfield  {title} {\bibinfo
  {title} {Results from phase 1 of the haystac microwave cavity axion
  experiment},\ }\href {https://doi.org/10.1103/PhysRevD.97.092001} {\bibfield
  {journal} {\bibinfo  {journal} {Phys. Rev. D}\ }\textbf {\bibinfo {volume}
  {97}},\ \bibinfo {pages} {092001} (\bibinfo {year} {2018})}\BibitemShut
  {NoStop}%
\bibitem [{\citenamefont {Backes}\ \emph {et~al.}(2021)\citenamefont {Backes},
  \citenamefont {Palken}, \citenamefont {Kenany}, \citenamefont {Brubaker},
  \citenamefont {Cahn}, \citenamefont {Droster}, \citenamefont {Hilton},
  \citenamefont {Ghosh}, \citenamefont {Jackson}, \citenamefont {Lamoreaux}
  \emph {et~al.}}]{HAYSTAC2021}%
  \BibitemOpen
  \bibfield  {author} {\bibinfo {author} {\bibfnamefont {K.~M.}\ \bibnamefont
  {Backes}}, \bibinfo {author} {\bibfnamefont {D.~A.}\ \bibnamefont {Palken}},
  \bibinfo {author} {\bibfnamefont {S.~A.}\ \bibnamefont {Kenany}}, \bibinfo
  {author} {\bibfnamefont {B.~M.}\ \bibnamefont {Brubaker}}, \bibinfo {author}
  {\bibfnamefont {S.~B.}\ \bibnamefont {Cahn}}, \bibinfo {author}
  {\bibfnamefont {A.}~\bibnamefont {Droster}}, \bibinfo {author} {\bibfnamefont
  {G.~C.}\ \bibnamefont {Hilton}}, \bibinfo {author} {\bibfnamefont
  {S.}~\bibnamefont {Ghosh}}, \bibinfo {author} {\bibfnamefont
  {H.}~\bibnamefont {Jackson}}, \bibinfo {author} {\bibfnamefont {S.~K.}\
  \bibnamefont {Lamoreaux}}, \emph {et~al.},\ }\bibfield  {title} {\bibinfo
  {title} {A quantum enhanced search for dark matter axions},\ }\href
  {https://doi.org/10.1038/s41586-021-03226-7} {\bibfield  {journal} {\bibinfo
  {journal} {Nature}\ }\textbf {\bibinfo {volume} {590}},\ \bibinfo {pages}
  {238–242} (\bibinfo {year} {2021})}\BibitemShut {NoStop}%
\bibitem [{\citenamefont {Sikivie}(1983)}]{Sikivie1983}%
  \BibitemOpen
  \bibfield  {author} {\bibinfo {author} {\bibfnamefont {P.}~\bibnamefont
  {Sikivie}},\ }\bibfield  {title} {\bibinfo {title} {Experimental tests of the
  ``invisible'' axion},\ }\href {https://doi.org/10.1103/PhysRevLett.51.1415}
  {\bibfield  {journal} {\bibinfo  {journal} {Phys. Rev. Lett.}\ }\textbf
  {\bibinfo {volume} {51}},\ \bibinfo {pages} {1415} (\bibinfo {year}
  {1983})}\BibitemShut {NoStop}%
\bibitem [{\citenamefont {Sikivie}(1985)}]{Sikivie1985}%
  \BibitemOpen
  \bibfield  {author} {\bibinfo {author} {\bibfnamefont {P.}~\bibnamefont
  {Sikivie}},\ }\bibfield  {title} {\bibinfo {title} {Detection rates for
  ``invisible''-axion searches},\ }\href
  {https://doi.org/10.1103/PhysRevD.32.2988} {\bibfield  {journal} {\bibinfo
  {journal} {Phys. Rev. D}\ }\textbf {\bibinfo {volume} {32}},\ \bibinfo
  {pages} {2988} (\bibinfo {year} {1985})}\BibitemShut {NoStop}%
\bibitem [{\citenamefont {Tada}\ \emph {et~al.}(1999)\citenamefont {Tada},
  \citenamefont {Kishimoto}, \citenamefont {Kominato}, \citenamefont {Shibata},
  \citenamefont {Funahashi}, \citenamefont {Yamamoto}, \citenamefont
  {Masaike},\ and\ \citenamefont {Matsuki}}]{tada1999}%
  \BibitemOpen
  \bibfield  {author} {\bibinfo {author} {\bibfnamefont {M.}~\bibnamefont
  {Tada}}, \bibinfo {author} {\bibfnamefont {Y.}~\bibnamefont {Kishimoto}},
  \bibinfo {author} {\bibfnamefont {K.}~\bibnamefont {Kominato}}, \bibinfo
  {author} {\bibfnamefont {M.}~\bibnamefont {Shibata}}, \bibinfo {author}
  {\bibfnamefont {H.}~\bibnamefont {Funahashi}}, \bibinfo {author}
  {\bibfnamefont {K.}~\bibnamefont {Yamamoto}}, \bibinfo {author}
  {\bibfnamefont {A.}~\bibnamefont {Masaike}},\ and\ \bibinfo {author}
  {\bibfnamefont {S.}~\bibnamefont {Matsuki}},\ }\bibfield  {title} {\bibinfo
  {title} {{CARRACK II} — a new large-scale experiment to search for axions
  with {R}ydberg-atom cavity detector},\ }\href
  {https://doi.org/https://doi.org/10.1016/S0920-5632(98)00519-2} {\bibfield
  {journal} {\bibinfo  {journal} {Nuclear Physics B - Proceedings Supplements}\
  }\textbf {\bibinfo {volume} {72}},\ \bibinfo {pages} {164} (\bibinfo {year}
  {1999})},\ \bibinfo {note} {proceedings of the 5th IFT Workshop on
  Axions}\BibitemShut {NoStop}%
\bibitem [{\citenamefont {Yamamoto}\ \emph {et~al.}(2001)\citenamefont
  {Yamamoto}, \citenamefont {Tada}, \citenamefont {Kishimoto}, \citenamefont
  {Shibata}, \citenamefont {Kominato}, \citenamefont {Ooishi}, \citenamefont
  {Yamada}, \citenamefont {Saida}, \citenamefont {Funahashi}, \citenamefont
  {Masaike},\ and\ \citenamefont {Matsuki}}]{Yamamoto2001}%
  \BibitemOpen
  \bibfield  {author} {\bibinfo {author} {\bibfnamefont {K.}~\bibnamefont
  {Yamamoto}}, \bibinfo {author} {\bibfnamefont {M.}~\bibnamefont {Tada}},
  \bibinfo {author} {\bibfnamefont {Y.}~\bibnamefont {Kishimoto}}, \bibinfo
  {author} {\bibfnamefont {M.}~\bibnamefont {Shibata}}, \bibinfo {author}
  {\bibfnamefont {K.}~\bibnamefont {Kominato}}, \bibinfo {author}
  {\bibfnamefont {T.}~\bibnamefont {Ooishi}}, \bibinfo {author} {\bibfnamefont
  {S.}~\bibnamefont {Yamada}}, \bibinfo {author} {\bibfnamefont
  {T.}~\bibnamefont {Saida}}, \bibinfo {author} {\bibfnamefont
  {H.}~\bibnamefont {Funahashi}}, \bibinfo {author} {\bibfnamefont
  {A.}~\bibnamefont {Masaike}},\ and\ \bibinfo {author} {\bibfnamefont
  {S.}~\bibnamefont {Matsuki}},\ }\bibfield  {title} {\bibinfo {title} {The
  rydberg-atom-cavity axion search},\ }in\ \href@noop {} {\emph {\bibinfo
  {booktitle} {Dark Matter in Astro- and Particle Physics}}},\ \bibinfo
  {editor} {edited by\ \bibinfo {editor} {\bibfnamefont {H.~V.}\ \bibnamefont
  {Klapdor-Kleingrothaus}}}\ (\bibinfo  {publisher} {Springer Berlin
  Heidelberg},\ \bibinfo {address} {Berlin, Heidelberg},\ \bibinfo {year}
  {2001})\ pp.\ \bibinfo {pages} {638--645}\BibitemShut {NoStop}%
\bibitem [{\citenamefont {Lamoreaux}\ \emph {et~al.}(2013)\citenamefont
  {Lamoreaux}, \citenamefont {van Bibber}, \citenamefont {Lehnert},\ and\
  \citenamefont {Carosi}}]{Lamoreaux2013}%
  \BibitemOpen
  \bibfield  {author} {\bibinfo {author} {\bibfnamefont {S.~K.}\ \bibnamefont
  {Lamoreaux}}, \bibinfo {author} {\bibfnamefont {K.~A.}\ \bibnamefont {van
  Bibber}}, \bibinfo {author} {\bibfnamefont {K.~W.}\ \bibnamefont {Lehnert}},\
  and\ \bibinfo {author} {\bibfnamefont {G.}~\bibnamefont {Carosi}},\
  }\bibfield  {title} {\bibinfo {title} {Analysis of single-photon and linear
  amplifier detectors for microwave cavity dark matter axion searches},\ }\href
  {http://dx.doi.org/10.1103/PhysRevD.88.035020} {\bibfield  {journal}
  {\bibinfo  {journal} {Physical Review D}\ }\textbf {\bibinfo {volume} {88}}
  (\bibinfo {year} {2013})}\BibitemShut {NoStop}%
\bibitem [{\citenamefont {Tada}\ \emph {et~al.}(2006)\citenamefont {Tada},
  \citenamefont {Kishimoto}, \citenamefont {Kominato}, \citenamefont {Shibata},
  \citenamefont {Yamada}, \citenamefont {Haseyama}, \citenamefont {Ogawa},
  \citenamefont {Funahashi}, \citenamefont {Yamamoto},\ and\ \citenamefont
  {Matsuki}}]{TADA2006}%
  \BibitemOpen
  \bibfield  {author} {\bibinfo {author} {\bibfnamefont {M.}~\bibnamefont
  {Tada}}, \bibinfo {author} {\bibfnamefont {Y.}~\bibnamefont {Kishimoto}},
  \bibinfo {author} {\bibfnamefont {K.}~\bibnamefont {Kominato}}, \bibinfo
  {author} {\bibfnamefont {M.}~\bibnamefont {Shibata}}, \bibinfo {author}
  {\bibfnamefont {S.}~\bibnamefont {Yamada}}, \bibinfo {author} {\bibfnamefont
  {T.}~\bibnamefont {Haseyama}}, \bibinfo {author} {\bibfnamefont
  {I.}~\bibnamefont {Ogawa}}, \bibinfo {author} {\bibfnamefont
  {H.}~\bibnamefont {Funahashi}}, \bibinfo {author} {\bibfnamefont
  {K.}~\bibnamefont {Yamamoto}},\ and\ \bibinfo {author} {\bibfnamefont
  {S.}~\bibnamefont {Matsuki}},\ }\bibfield  {title} {\bibinfo {title}
  {Single-photon detection of microwave blackbody radiations in a
  low-temperature resonant-cavity with high {Rydberg} atoms},\ }\href
  {https://doi.org/https://doi.org/10.1016/j.physleta.2005.09.066} {\bibfield
  {journal} {\bibinfo  {journal} {Physics Letters A}\ }\textbf {\bibinfo
  {volume} {349}},\ \bibinfo {pages} {488} (\bibinfo {year}
  {2006})}\BibitemShut {NoStop}%
\bibitem [{\citenamefont {Ballesteros}\ \emph {et~al.}(2017)\citenamefont
  {Ballesteros}, \citenamefont {Redondo}, \citenamefont {Ringwald},\ and\
  \citenamefont {Tamarit}}]{Ballesteros2017}%
  \BibitemOpen
  \bibfield  {author} {\bibinfo {author} {\bibfnamefont {G.}~\bibnamefont
  {Ballesteros}}, \bibinfo {author} {\bibfnamefont {J.}~\bibnamefont
  {Redondo}}, \bibinfo {author} {\bibfnamefont {A.}~\bibnamefont {Ringwald}},\
  and\ \bibinfo {author} {\bibfnamefont {C.}~\bibnamefont {Tamarit}},\
  }\bibfield  {title} {\bibinfo {title} {Unifying inflation with the axion,
  dark matter, baryogenesis, and the seesaw mechanism},\ }\href
  {https://doi.org/10.1103/PhysRevLett.118.071802} {\bibfield  {journal}
  {\bibinfo  {journal} {Phys. Rev. Lett.}\ }\textbf {\bibinfo {volume} {118}},\
  \bibinfo {pages} {071802} (\bibinfo {year} {2017})}\BibitemShut {NoStop}%
\bibitem [{\citenamefont {Gorghetto}\ \emph {et~al.}(2021)\citenamefont
  {Gorghetto}, \citenamefont {Hardy},\ and\ \citenamefont
  {Villadoro}}]{Gorghetto2021}%
  \BibitemOpen
  \bibfield  {author} {\bibinfo {author} {\bibfnamefont {M.}~\bibnamefont
  {Gorghetto}}, \bibinfo {author} {\bibfnamefont {E.}~\bibnamefont {Hardy}},\
  and\ \bibinfo {author} {\bibfnamefont {G.}~\bibnamefont {Villadoro}},\
  }\bibfield  {title} {\bibinfo {title} {{More Axions from Strings}},\ }\href
  {https://doi.org/10.21468/SciPostPhys.10.2.050} {\bibfield  {journal}
  {\bibinfo  {journal} {SciPost Phys.}\ }\textbf {\bibinfo {volume} {10}},\
  \bibinfo {pages} {50} (\bibinfo {year} {2021})}\BibitemShut {NoStop}%
\bibitem [{\citenamefont {Buschmann}\ \emph {et~al.}(2021)\citenamefont
  {Buschmann}, \citenamefont {Foster}, \citenamefont {Hook}, \citenamefont
  {Peterson}, \citenamefont {Willcox}, \citenamefont {Zhang},\ and\
  \citenamefont {Safdi}}]{buschmann2021}%
  \BibitemOpen
  \bibfield  {author} {\bibinfo {author} {\bibfnamefont {M.}~\bibnamefont
  {Buschmann}}, \bibinfo {author} {\bibfnamefont {J.~W.}\ \bibnamefont
  {Foster}}, \bibinfo {author} {\bibfnamefont {A.}~\bibnamefont {Hook}},
  \bibinfo {author} {\bibfnamefont {A.}~\bibnamefont {Peterson}}, \bibinfo
  {author} {\bibfnamefont {D.~E.}\ \bibnamefont {Willcox}}, \bibinfo {author}
  {\bibfnamefont {W.}~\bibnamefont {Zhang}},\ and\ \bibinfo {author}
  {\bibfnamefont {B.~R.}\ \bibnamefont {Safdi}},\ }\href@noop {} {\bibinfo
  {title} {Dark matter from axion strings with adaptive mesh refinement}}
  (\bibinfo {year} {2021}),\ \Eprint {https://arxiv.org/abs/2108.05368}
  {arXiv:2108.05368 [hep-ph]} \BibitemShut {NoStop}%
\bibitem [{\citenamefont {Lorenzen}\ \emph {et~al.}(1981)\citenamefont
  {Lorenzen}, \citenamefont {Niemax},\ and\ \citenamefont
  {Pendrill}}]{Lorenzen1981}%
  \BibitemOpen
  \bibfield  {author} {\bibinfo {author} {\bibfnamefont {C.-J.}\ \bibnamefont
  {Lorenzen}}, \bibinfo {author} {\bibfnamefont {K.}~\bibnamefont {Niemax}},\
  and\ \bibinfo {author} {\bibfnamefont {L.}~\bibnamefont {Pendrill}},\
  }\bibfield  {title} {\bibinfo {title} {Precise measurements of {$^{39}$K}
  n{S} and n{D} energy levels with an evaluated wavemeter},\ }\href
  {https://doi.org/https://doi.org/10.1016/0030-4018(81)90225-X} {\bibfield
  {journal} {\bibinfo  {journal} {Optics Communications}\ }\textbf {\bibinfo
  {volume} {39}},\ \bibinfo {pages} {370} (\bibinfo {year} {1981})}\BibitemShut
  {NoStop}%
\bibitem [{\citenamefont {Lorenzen}\ and\ \citenamefont
  {Niemax}(1983)}]{Lorenzen1983}%
  \BibitemOpen
  \bibfield  {author} {\bibinfo {author} {\bibfnamefont {C.-J.}\ \bibnamefont
  {Lorenzen}}\ and\ \bibinfo {author} {\bibfnamefont {K.}~\bibnamefont
  {Niemax}},\ }\bibfield  {title} {\bibinfo {title} {Quantum defects of the
  {$n^2P_{1/2,3/2}$} levels in {$^{39}$K I} and {$^{85}$Rb I}},\ }\href
  {https://doi.org/10.1088/0031-8949/27/4/012} {\bibfield  {journal} {\bibinfo
  {journal} {Physica Scripta}\ }\textbf {\bibinfo {volume} {27}},\ \bibinfo
  {pages} {300} (\bibinfo {year} {1983})}\BibitemShut {NoStop}%
\bibitem [{\citenamefont {Thompson}\ \emph {et~al.}(1983)\citenamefont
  {Thompson}, \citenamefont {O'Sullivan}, \citenamefont {Stoicheff},\ and\
  \citenamefont {Xu}}]{Thompson1983}%
  \BibitemOpen
  \bibfield  {author} {\bibinfo {author} {\bibfnamefont {D.~C.}\ \bibnamefont
  {Thompson}}, \bibinfo {author} {\bibfnamefont {M.~S.}\ \bibnamefont
  {O'Sullivan}}, \bibinfo {author} {\bibfnamefont {B.~P.}\ \bibnamefont
  {Stoicheff}},\ and\ \bibinfo {author} {\bibfnamefont {G.-X.}\ \bibnamefont
  {Xu}},\ }\bibfield  {title} {\bibinfo {title} {Doppler-free two-photon
  absorption spectrum of potassium},\ }\href {https://doi.org/10.1139/p83-117}
  {\bibfield  {journal} {\bibinfo  {journal} {Canadian Journal of Physics}\
  }\textbf {\bibinfo {volume} {61}},\ \bibinfo {pages} {949} (\bibinfo {year}
  {1983})}\BibitemShut {NoStop}%
\bibitem [{\citenamefont {Chen}\ \emph {et~al.}(2020)\citenamefont {Chen},
  \citenamefont {Chang}, \citenamefont {Huang}, \citenamefont {Shukla},
  \citenamefont {Huang}, \citenamefont {Suen}, \citenamefont {Kuan},
  \citenamefont {Shy},\ and\ \citenamefont {Liu}}]{Chen2020}%
  \BibitemOpen
  \bibfield  {author} {\bibinfo {author} {\bibfnamefont {T.-L.}\ \bibnamefont
  {Chen}}, \bibinfo {author} {\bibfnamefont {S.-Y.}\ \bibnamefont {Chang}},
  \bibinfo {author} {\bibfnamefont {Y.-J.}\ \bibnamefont {Huang}}, \bibinfo
  {author} {\bibfnamefont {K.}~\bibnamefont {Shukla}}, \bibinfo {author}
  {\bibfnamefont {Y.-C.}\ \bibnamefont {Huang}}, \bibinfo {author}
  {\bibfnamefont {T.-H.}\ \bibnamefont {Suen}}, \bibinfo {author}
  {\bibfnamefont {T.-Y.}\ \bibnamefont {Kuan}}, \bibinfo {author}
  {\bibfnamefont {J.-T.}\ \bibnamefont {Shy}},\ and\ \bibinfo {author}
  {\bibfnamefont {Y.-W.}\ \bibnamefont {Liu}},\ }\bibfield  {title} {\bibinfo
  {title} {Inverted-ladder-type optical excitation of potassium rydberg states
  with hot and cold ensembles},\ }\href
  {https://doi.org/10.1103/PhysRevA.101.052507} {\bibfield  {journal} {\bibinfo
   {journal} {Phys. Rev. A}\ }\textbf {\bibinfo {volume} {101}},\ \bibinfo
  {pages} {052507} (\bibinfo {year} {2020})}\BibitemShut {NoStop}%
\bibitem [{\citenamefont {Boller}\ \emph {et~al.}(1991)\citenamefont {Boller},
  \citenamefont {Imamo\ifmmode~\breve{g}\else \u{g}\fi{}lu},\ and\
  \citenamefont {Harris}}]{Boller1991}%
  \BibitemOpen
  \bibfield  {author} {\bibinfo {author} {\bibfnamefont {K.-J.}\ \bibnamefont
  {Boller}}, \bibinfo {author} {\bibfnamefont {A.}~\bibnamefont
  {Imamo\ifmmode~\breve{g}\else \u{g}\fi{}lu}},\ and\ \bibinfo {author}
  {\bibfnamefont {S.~E.}\ \bibnamefont {Harris}},\ }\bibfield  {title}
  {\bibinfo {title} {Observation of electromagnetically induced transparency},\
  }\href {https://doi.org/10.1103/PhysRevLett.66.2593} {\bibfield  {journal}
  {\bibinfo  {journal} {Phys. Rev. Lett.}\ }\textbf {\bibinfo {volume} {66}},\
  \bibinfo {pages} {2593} (\bibinfo {year} {1991})}\BibitemShut {NoStop}%
\bibitem [{\citenamefont {Mohapatra}\ \emph {et~al.}(2007)\citenamefont
  {Mohapatra}, \citenamefont {Jackson},\ and\ \citenamefont
  {Adams}}]{Mohapatra2007}%
  \BibitemOpen
  \bibfield  {author} {\bibinfo {author} {\bibfnamefont {A.~K.}\ \bibnamefont
  {Mohapatra}}, \bibinfo {author} {\bibfnamefont {T.~R.}\ \bibnamefont
  {Jackson}},\ and\ \bibinfo {author} {\bibfnamefont {C.~S.}\ \bibnamefont
  {Adams}},\ }\bibfield  {title} {\bibinfo {title} {Coherent optical detection
  of highly excited rydberg states using electromagnetically induced
  transparency},\ }\href {https://doi.org/10.1103/PhysRevLett.98.113003}
  {\bibfield  {journal} {\bibinfo  {journal} {Phys. Rev. Lett.}\ }\textbf
  {\bibinfo {volume} {98}},\ \bibinfo {pages} {113003} (\bibinfo {year}
  {2007})}\BibitemShut {NoStop}%
\bibitem [{\citenamefont {Xu}\ and\ \citenamefont {DeMarco}(2016)}]{Xu2016}%
  \BibitemOpen
  \bibfield  {author} {\bibinfo {author} {\bibfnamefont {W.}~\bibnamefont
  {Xu}}\ and\ \bibinfo {author} {\bibfnamefont {B.}~\bibnamefont {DeMarco}},\
  }\bibfield  {title} {\bibinfo {title} {Velocity-selective
  electromagnetically-induced-transparency measurements of potassium rydberg
  states},\ }\href {https://doi.org/10.1103/PhysRevA.93.011801} {\bibfield
  {journal} {\bibinfo  {journal} {Phys. Rev. A}\ }\textbf {\bibinfo {volume}
  {93}},\ \bibinfo {pages} {011801} (\bibinfo {year} {2016})}\BibitemShut
  {NoStop}%
\bibitem [{\citenamefont {Saffman}\ \emph {et~al.}(2010)\citenamefont
  {Saffman}, \citenamefont {Walker},\ and\ \citenamefont
  {M\o{}lmer}}]{Saffman2010}%
  \BibitemOpen
  \bibfield  {author} {\bibinfo {author} {\bibfnamefont {M.}~\bibnamefont
  {Saffman}}, \bibinfo {author} {\bibfnamefont {T.~G.}\ \bibnamefont
  {Walker}},\ and\ \bibinfo {author} {\bibfnamefont {K.}~\bibnamefont
  {M\o{}lmer}},\ }\bibfield  {title} {\bibinfo {title} {Quantum information
  with {Rydberg} atoms},\ }\href {https://doi.org/10.1103/RevModPhys.82.2313}
  {\bibfield  {journal} {\bibinfo  {journal} {Rev. Mod. Phys.}\ }\textbf
  {\bibinfo {volume} {82}},\ \bibinfo {pages} {2313} (\bibinfo {year}
  {2010})}\BibitemShut {NoStop}%
\bibitem [{\citenamefont {Pupillo}\ \emph {et~al.}(2010)\citenamefont
  {Pupillo}, \citenamefont {Micheli}, \citenamefont {Boninsegni}, \citenamefont
  {Lesanovsky},\ and\ \citenamefont {Zoller}}]{Pupillo2010}%
  \BibitemOpen
  \bibfield  {author} {\bibinfo {author} {\bibfnamefont {G.}~\bibnamefont
  {Pupillo}}, \bibinfo {author} {\bibfnamefont {A.}~\bibnamefont {Micheli}},
  \bibinfo {author} {\bibfnamefont {M.}~\bibnamefont {Boninsegni}}, \bibinfo
  {author} {\bibfnamefont {I.}~\bibnamefont {Lesanovsky}},\ and\ \bibinfo
  {author} {\bibfnamefont {P.}~\bibnamefont {Zoller}},\ }\bibfield  {title}
  {\bibinfo {title} {Strongly correlated gases of {Rydberg}-dressed atoms:
  Quantum and classical dynamics},\ }\href
  {https://doi.org/10.1103/PhysRevLett.104.223002} {\bibfield  {journal}
  {\bibinfo  {journal} {Phys. Rev. Lett.}\ }\textbf {\bibinfo {volume} {104}},\
  \bibinfo {pages} {223002} (\bibinfo {year} {2010})}\BibitemShut {NoStop}%
\bibitem [{\citenamefont {Boon}\ \emph {et~al.}(1999)\citenamefont {Boon},
  \citenamefont {Zekou}, \citenamefont {McGloin},\ and\ \citenamefont
  {Dunn}}]{Boon1999}%
  \BibitemOpen
  \bibfield  {author} {\bibinfo {author} {\bibfnamefont {J.~R.}\ \bibnamefont
  {Boon}}, \bibinfo {author} {\bibfnamefont {E.}~\bibnamefont {Zekou}},
  \bibinfo {author} {\bibfnamefont {D.}~\bibnamefont {McGloin}},\ and\ \bibinfo
  {author} {\bibfnamefont {M.~H.}\ \bibnamefont {Dunn}},\ }\bibfield  {title}
  {\bibinfo {title} {Comparison of wavelength dependence in cascade-,
  \ensuremath{\Lambda}-, and {Vee}-type schemes for electromagnetically induced
  transparency},\ }\href {https://doi.org/10.1103/PhysRevA.59.4675} {\bibfield
  {journal} {\bibinfo  {journal} {Phys. Rev. A}\ }\textbf {\bibinfo {volume}
  {59}},\ \bibinfo {pages} {4675} (\bibinfo {year} {1999})}\BibitemShut
  {NoStop}%
\bibitem [{\citenamefont {Urvoy}\ \emph {et~al.}(2013)\citenamefont {Urvoy},
  \citenamefont {Carr}, \citenamefont {Ritter}, \citenamefont {Adams},
  \citenamefont {Weatherill},\ and\ \citenamefont {Löw}}]{Urvoy2013}%
  \BibitemOpen
  \bibfield  {author} {\bibinfo {author} {\bibfnamefont {A.}~\bibnamefont
  {Urvoy}}, \bibinfo {author} {\bibfnamefont {C.}~\bibnamefont {Carr}},
  \bibinfo {author} {\bibfnamefont {R.}~\bibnamefont {Ritter}}, \bibinfo
  {author} {\bibfnamefont {C.~S.}\ \bibnamefont {Adams}}, \bibinfo {author}
  {\bibfnamefont {K.~J.}\ \bibnamefont {Weatherill}},\ and\ \bibinfo {author}
  {\bibfnamefont {R.}~\bibnamefont {Löw}},\ }\bibfield  {title} {\bibinfo
  {title} {Optical coherences and wavelength mismatch in ladder systems},\
  }\href {https://doi.org/10.1088/0953-4075/46/24/245001} {\bibfield  {journal}
  {\bibinfo  {journal} {J. Phys. B: At. Mol. Opt. Phys.}\ }\textbf {\bibinfo
  {volume} {46}},\ \bibinfo {pages} {245001} (\bibinfo {year}
  {2013})}\BibitemShut {NoStop}%
\bibitem [{\citenamefont {Sansonetti}(2008)}]{Sansonetti}%
  \BibitemOpen
  \bibfield  {author} {\bibinfo {author} {\bibfnamefont {J.~E.}\ \bibnamefont
  {Sansonetti}},\ }\bibfield  {title} {\bibinfo {title} {Wavelengths,
  transition probabilities, and energy levels for the spectra of potassium
  ({KI} through {KXIX})},\ }\href@noop {} {\bibfield  {journal} {\bibinfo
  {journal} {Journal of Physical and Chemical Reference Data}\ }\textbf
  {\bibinfo {volume} {37}},\ \bibinfo {pages} {7} (\bibinfo {year}
  {2008})}\BibitemShut {NoStop}%
\bibitem [{\citenamefont {Gallagher}(1994)}]{gallagher1994}%
  \BibitemOpen
  \bibfield  {author} {\bibinfo {author} {\bibfnamefont {T.~F.}\ \bibnamefont
  {Gallagher}},\ }\href {https://doi.org/10.1017/CBO9780511524530} {\emph
  {\bibinfo {title} {{Rydberg} Atoms}}},\ Cambridge Monographs on Atomic,
  Molecular and Chemical Physics\ (\bibinfo  {publisher} {Cambridge University
  Press},\ \bibinfo {year} {1994})\BibitemShut {NoStop}%
\bibitem [{\citenamefont {Mohr}\ \emph {et~al.}(2016)\citenamefont {Mohr},
  \citenamefont {Newell},\ and\ \citenamefont {Taylor}}]{NISTCODATA}%
  \BibitemOpen
  \bibfield  {author} {\bibinfo {author} {\bibfnamefont {P.~J.}\ \bibnamefont
  {Mohr}}, \bibinfo {author} {\bibfnamefont {D.~B.}\ \bibnamefont {Newell}},\
  and\ \bibinfo {author} {\bibfnamefont {B.~N.}\ \bibnamefont {Taylor}},\
  }\bibfield  {title} {\bibinfo {title} {{CODATA} recommended values of the
  fundamental physical constants: 2014},\ }\href
  {https://doi.org/10.1103/RevModPhys.88.035009} {\bibfield  {journal}
  {\bibinfo  {journal} {Rev. Mod. Phys.}\ }\textbf {\bibinfo {volume} {88}},\
  \bibinfo {pages} {035009} (\bibinfo {year} {2016})}\BibitemShut {NoStop}%
\bibitem [{\citenamefont {Coursey}\ \emph {et~al.}(2015)\citenamefont
  {Coursey}, \citenamefont {Schwab}, \citenamefont {Tsai},\ and\ \citenamefont
  {Dragoset}}]{AtomicWeight}%
  \BibitemOpen
  \bibfield  {author} {\bibinfo {author} {\bibfnamefont {J.}~\bibnamefont
  {Coursey}}, \bibinfo {author} {\bibfnamefont {D.}~\bibnamefont {Schwab}},
  \bibinfo {author} {\bibfnamefont {J.}~\bibnamefont {Tsai}},\ and\ \bibinfo
  {author} {\bibfnamefont {R.}~\bibnamefont {Dragoset}},\ }\href@noop {}
  {}\bibinfo {howpublished} {Atomic Weights and Isotopic Compositions (version
  4.1), [Online]. Available: {\href{http://physics.nist.gov/Comp}{
  http://physics.nist.gov/Comp}} [2022, Jan 11]. National Institute of
  Standards and Technology, Gaithersburg, MD} (\bibinfo {year}
  {2015})\BibitemShut {NoStop}%
\bibitem [{\citenamefont {Peper}\ \emph {et~al.}(2019)\citenamefont {Peper},
  \citenamefont {Helmrich}, \citenamefont {Butscher}, \citenamefont {Agner},
  \citenamefont {Schmutz}, \citenamefont {Merkt},\ and\ \citenamefont
  {Deiglmayr}}]{Peper2019}%
  \BibitemOpen
  \bibfield  {author} {\bibinfo {author} {\bibfnamefont {M.}~\bibnamefont
  {Peper}}, \bibinfo {author} {\bibfnamefont {F.}~\bibnamefont {Helmrich}},
  \bibinfo {author} {\bibfnamefont {J.}~\bibnamefont {Butscher}}, \bibinfo
  {author} {\bibfnamefont {J.~A.}\ \bibnamefont {Agner}}, \bibinfo {author}
  {\bibfnamefont {H.}~\bibnamefont {Schmutz}}, \bibinfo {author} {\bibfnamefont
  {F.}~\bibnamefont {Merkt}},\ and\ \bibinfo {author} {\bibfnamefont
  {J.}~\bibnamefont {Deiglmayr}},\ }\bibfield  {title} {\bibinfo {title}
  {Precision measurement of the ionization energy and quantum defects of
  $^{39}\mathrm{K}$ {I}},\ }\href {https://doi.org/10.1103/PhysRevA.100.012501}
  {\bibfield  {journal} {\bibinfo  {journal} {Phys. Rev. A}\ }\textbf {\bibinfo
  {volume} {100}},\ \bibinfo {pages} {012501} (\bibinfo {year}
  {2019})}\BibitemShut {NoStop}%
\bibitem [{\citenamefont {Martin}(1980)}]{Martin1980}%
  \BibitemOpen
  \bibfield  {author} {\bibinfo {author} {\bibfnamefont {W.~C.}\ \bibnamefont
  {Martin}},\ }\bibfield  {title} {\bibinfo {title} {Series formulas for the
  spectrum of atomic sodium ({NaI})},\ }\href
  {https://doi.org/10.1364/JOSA.70.000784} {\bibfield  {journal} {\bibinfo
  {journal} {J. Opt. Soc. Am.}\ }\textbf {\bibinfo {volume} {70}},\ \bibinfo
  {pages} {784} (\bibinfo {year} {1980})}\BibitemShut {NoStop}%
\bibitem [{\citenamefont {Drake}\ and\ \citenamefont
  {Swainson}(1991)}]{Drake1991}%
  \BibitemOpen
  \bibfield  {author} {\bibinfo {author} {\bibfnamefont {G.~W.~F.}\
  \bibnamefont {Drake}}\ and\ \bibinfo {author} {\bibfnamefont {R.~A.}\
  \bibnamefont {Swainson}},\ }\bibfield  {title} {\bibinfo {title} {Quantum
  defects and the $1/n$ dependence of {R}ydberg energies: Second-order
  polarization effects},\ }\href {https://doi.org/10.1103/PhysRevA.44.5448}
  {\bibfield  {journal} {\bibinfo  {journal} {Phys. Rev. A}\ }\textbf {\bibinfo
  {volume} {44}},\ \bibinfo {pages} {5448} (\bibinfo {year}
  {1991})}\BibitemShut {NoStop}%
\bibitem [{\citenamefont {Kramida}\ \emph {et~al.}(2020)\citenamefont
  {Kramida}, \citenamefont {{Yu.~Ralchenko}}, \citenamefont {Reader},\ and\
  \citenamefont {{and NIST ASD Team}}}]{NIST_ASD}%
  \BibitemOpen
  \bibfield  {author} {\bibinfo {author} {\bibfnamefont {A.}~\bibnamefont
  {Kramida}}, \bibinfo {author} {\bibnamefont {{Yu.~Ralchenko}}}, \bibinfo
  {author} {\bibfnamefont {J.}~\bibnamefont {Reader}},\ and\ \bibinfo {author}
  {\bibnamefont {{and NIST ASD Team}}},\ }\href@noop {} {}\bibinfo
  {howpublished} {{NIST Atomic Spectra Database (ver. 5.8), [Online].
  Available:
  {\href{https://physics.nist.gov/asd}{https://physics.nist.gov/asd}} [2021,
  August 23]. National Institute of Standards and Technology, Gaithersburg,
  MD}} (\bibinfo {year} {2020})\BibitemShut {NoStop}%
\bibitem [{\citenamefont {Šibalić}\ \emph {et~al.}(2017)\citenamefont
  {Šibalić}, \citenamefont {Pritchard}, \citenamefont {Adams},\ and\
  \citenamefont {Weatherill}}]{ARC}%
  \BibitemOpen
  \bibfield  {author} {\bibinfo {author} {\bibfnamefont {N.}~\bibnamefont
  {Šibalić}}, \bibinfo {author} {\bibfnamefont {J.}~\bibnamefont
  {Pritchard}}, \bibinfo {author} {\bibfnamefont {C.}~\bibnamefont {Adams}},\
  and\ \bibinfo {author} {\bibfnamefont {K.}~\bibnamefont {Weatherill}},\
  }\bibfield  {title} {\bibinfo {title} {Arc: An open-source library for
  calculating properties of alkali {R}ydberg atoms},\ }\href
  {https://doi.org/https://doi.org/10.1016/j.cpc.2017.06.015} {\bibfield
  {journal} {\bibinfo  {journal} {Computer Physics Communications}\ }\textbf
  {\bibinfo {volume} {220}},\ \bibinfo {pages} {319} (\bibinfo {year}
  {2017})}\BibitemShut {NoStop}%
\bibitem [{\citenamefont {Signoles}\ \emph {et~al.}(2017)\citenamefont
  {Signoles}, \citenamefont {Dietsche}, \citenamefont {Facon}, \citenamefont
  {Grosso}, \citenamefont {Haroche}, \citenamefont {Raimond}, \citenamefont
  {Brune},\ and\ \citenamefont {Gleyzes}}]{Signoles2017}%
  \BibitemOpen
  \bibfield  {author} {\bibinfo {author} {\bibfnamefont {A.}~\bibnamefont
  {Signoles}}, \bibinfo {author} {\bibfnamefont {E.~K.}\ \bibnamefont
  {Dietsche}}, \bibinfo {author} {\bibfnamefont {A.}~\bibnamefont {Facon}},
  \bibinfo {author} {\bibfnamefont {D.}~\bibnamefont {Grosso}}, \bibinfo
  {author} {\bibfnamefont {S.}~\bibnamefont {Haroche}}, \bibinfo {author}
  {\bibfnamefont {J.~M.}\ \bibnamefont {Raimond}}, \bibinfo {author}
  {\bibfnamefont {M.}~\bibnamefont {Brune}},\ and\ \bibinfo {author}
  {\bibfnamefont {S.}~\bibnamefont {Gleyzes}},\ }\bibfield  {title} {\bibinfo
  {title} {Coherent transfer between low-angular-momentum and circular
  {R}ydberg states},\ }\href {https://doi.org/10.1103/PhysRevLett.118.253603}
  {\bibfield  {journal} {\bibinfo  {journal} {Phys. Rev. Lett.}\ }\textbf
  {\bibinfo {volume} {118}},\ \bibinfo {pages} {253603} (\bibinfo {year}
  {2017})}\BibitemShut {NoStop}%
\end{thebibliography}%

\end{document}